\documentclass[fleqn,usenatbib]{mnras}

\usepackage{newtxtext,newtxmath}
\usepackage{adjustbox}
\usepackage{caption}

\usepackage[T1]{fontenc}

\DeclareRobustCommand{\VAN}[3]{#2}
\let\VANthebibliography\thebibliography
\def\thebibliography{\DeclareRobustCommand{\VAN}[3]{##3}\VANthebibliography}

\usepackage{graphicx}	
\usepackage{amsmath}	
\usepackage[normalem]{ulem}

\newcommand{\dave}{\textsf{DAVE }}
\newcommand{\eleanor}{\textsf{eleanor }}
\newcommand{\qlp}{\textsf{QLP }}
\newcommand{\autovetter}{\textsf{Autovetter }}
\newcommand{\robovetter}{\textsf{Robovetter }}
\newcommand{\astronet}{\textsf{Astronet }}
\newcommand{\astroramjet}{\textsf{astroramjet }}
\newcommand{\vespa}{\textsf{vespa }}
\newcommand{\triceratops}{\textsf{triceratops }}

\title[The TESS Triple-9 Catalog]{The TESS Triple 9 Catalog: 999 uniformly vetted exoplanet candidates}

\author[Cacciapuoti et al.]{
Luca Cacciapuoti,$^{1,2,3}$ \thanks{E-mail: luca.cacciapuoti@eso.org}
Veselin B. Kostov,$^{4,5}$
Marc Kuchner,$^{4}$
Elisa V. Quintana,$^{4}$
Knicole D. Col\'on,$^{4}$
\newauthor
Jonathan Brande,$^{6}$
Susan E. Mullally,$^{7}$
Quadry Chance,$^{8}$
Jessie L. Christiansen,$^{9}$
John P. Ahlers,$^{10}$
\newauthor
Marco Z. Di Fraia,$^{11,16}$
Hugo A. Durantini Luca,$^{12,16}$
Riccardo M. Ienco,$^{3,16}$
Francesco Gallo,$^{3,16}$
\newauthor
Lucas T. de Lima,$^{13,16}$
Michiharu Hyogo,$^{14,16}$
Marc Andrés-Carcasona,$^{15,16}$
Aline U. Fornear,$^{16}$
\newauthor
Julien S. de Lambilly,$^{16}$
Ryan Salik$^{16,17}$
John M. Yablonsky$^{16}$
and Shaun Wallace$^{18}$
Sovan Acharya,$^{16,19}$
\\
\\
$^{1}$European Southern Observatory, Karl-Schwarzschild-Strasse 2 D-85748 Garching bei Munchen, Germany\\
$^{2}$Fakultat fur Physik, Ludwig-Maximilians-Universitat Munchen, Scheinerstr. 1, 81679 Munchen, Germany\\
$^{3}$Department of Physics "Ettore Pancini", University of Naples Federico II, Naples, Italy\\
$^{4}$NASA Goddard Space Flight Center, Greenbelt, MD 20771, USA\\
$^{5}$SETI Institute,189 Bernardo Ave, Suite 200, Mountain View, CA 94043, USA\\
$^{6}$Department of Physics \& Astronomy, University of Kansas, 1082 Malott, 1251 Wescoe Hall Dr., Lawrence, KS 66045, USA\\
$^{7}$Space Telescope Science Institute, 3700 San Martin Drive, Baltimore, MD, 21218, USA\\
$^{8}$University of Florida, Gainesville, FL 32611\\
$^{9}$Infrared Processing and Analysis Center, Caltech, Pasadena CA 91125, USA\\
$^{10}$Exoplanets and Stellar Astrophysics Laboratory, Code 667, NASA Goddard Space Flight Center (GSFC), Greenbelt, MD 20771, USA\\
$^{11}$Oxford Dynamics, G19, Building R71, RAL, Didcot OX11 0QX\\
$^{12}$IATE-OAC, Universidad Nacional de Córdoba-CONICET, Laprida 854, X5000 BGR, Córdoba, Argentina\\
$^{13}$Geoscience Department, University of Aveiro. Campus Santiago, 3810-193, Aveiro\\
$^{14}$Meisei University, 2-1-1 Hodokubo, Hino, Tokyo 191-0042, Japan\\
$^{15}$Institut de Física d’Altes Energies (IFAE), Barcelona Institute of Science and Technology, E-08193 Barcelona, Spain\\
$^{16}$Citizen Scientist, Planet Patrol Collaboration\\
$^{17}$Staples High School, 70 North Ave, Westport, Connecticut 06880\\
$^{18}$Department of Computer Science Brown University\\
$^{19}$SA Citizen Science Group-Ignited Minds VIPNET Club\\
}

\date{Accepted XXX. Received YYY; in original form ZZZ}

\pubyear{2022}

\begin{document}
\label{firstpage}
\pagerange{\pageref{firstpage}--\pageref{lastpage}}
\maketitle

\begin{abstract}
    The Transiting Exoplanet Survey Satellite (TESS) has detected thousands of exoplanet candidates since 2018, most of which have yet to be confirmed. A key step in the confirmation process of these candidates is ruling out false positives through vetting. Vetting also eases the burden on follow-up observations, provides input for demographics studies, and facilitates training machine learning algorithms. Here we present the TESS Triple-9 (TT9) catalog -- a uniformly-vetted catalog containing dispositions for 999 exoplanet candidates listed on ExoFOP-TESS, known as TESS Objects of Interest (TOIs). The TT9 was produced using the Discovery And Vetting of Exoplanets pipeline, \dave, and utilizing the power of citizen science as part of the Planet Patrol project. More than 70\% of the TOIs listed in the TT9 pass our diagnostic tests, and are thus marked as true planetary candidates. We flagged 144 candidates as false positives, and identified 146 as potential false positives. At the time of writing, the TT9 catalog contains $\sim20\%$ of the entire ExoFOP-TESS TOIs list, demonstrates the synergy between automated tools and citizen science, and represents the first stage of our efforts to vet all TOIs. The \dave generated results are publicly available on ExoFOP-TESS.
\end{abstract}




\begingroup
\let\clearpage\relax

\endgroup
\newpage

\section{Introduction}
A plethora of ground- and space-based exoplanet-hunting efforts have contributed to the exponential growth of the number of discovered planets orbiting stars other than our Sun. Prior to 2010, exoplanet discoveries were mainly based on precise radial velocity curves, e.g. \citep{Mayor1995}, \citep{Marcy1998}, \citep{Lovis2005}, \citep{Naef2010}. A paradigm shift in exoplanet discoveries started with NASA's Kepler mission \citep{Borucki2008}, launched in 2009. This endeavor enabled the scientific community to gather a large exoplanet sample via the photometric transit technique. This mission alone contributed to the confirmation of more than 2,700 exoplanets and the discovery of as many candidate exoplanets still to be confirmed\footnote{\label{note1}https://exoplanetarchive.ipac.caltech.edu}. 
he wealth of planetary-like signals discovered by photometric surveys will likely grow and expand thanks to large archival datasets, and ongoing and future planned survey missions, e.g. TESS \citep{Ricker2015} and PLATO \citep{Rauer2021}. 

Indeed, since 2018, the Transiting Exoplanet Survey Satellite (TESS) space mission has been surveying most of the sky and searching for additional transiting exoplanets. TESS has already identified more than 5,000 TESS Object of Interest (TOIs), i.e. candidate exoplanets, and is expected to detect thousands more, e.g. \citep{Barclay_2018}. However, only 199 out of the $\sim5,000$ TOIs listed on ExoFOP-TESS have been statistically validated and/or confirmed at the time of writing, according to the NASA Exoplanet Archive. Given that non-planetary astrophysical sources (most notably eclipsing binary stars, see e.g. \cite{Ciardi2018}) and/or systematic effects can often mimic exoplanet transits, revealing the true nature of the remaining TOIs requires ruling out possible false positive scenarios and/or follow-up observations.

Ideally, the precision radial velocity (PRV) technique, e.g. \citep{Baranne1996}, \citep{Pepe2004}, is utilized to confirm the planetary nature of a specific candidate. This is indeed the method that was used to discover the first confirmed exoplanet orbiting a Sun-like star: 51 Pegasi b \citep{Mayor1995}. Specifically, the stellar spectrum is obtained at different epochs to detect periodic Doppler shifts of its absorption lines due to the orbit of the star around the center of mass of the star-planet system. The major downside of the PRV technique is that spectroscopic observations are time-consuming. To be able to detect a planet around a star with the PRV method, two conditions should apply. First, the star has to be sufficiently bright to allow observing its spectrum in a reasonable amount of time and with a sufficient signal-to-noise ratio (SNR). Secondly, planets with short orbital periods facilitate the detection since this allows obtaining complete cycle radial velocity curves in a reasonable amount of observing time. Given the number of exoplanets an all-sky survey like TESS is expected to discover \citep{Barclay_2018}, it is unfeasible to obtain PRV observations for every star with a transiting planetary candidate.

Thus, vetting procedures play a critical role in the analysis of exoplanet candidates. First, vetting is needed to ease the burden of ground-based follow-up facilities and to validate planets for which no PRV measurements can be obtained. Secondly, it is fundamental to refine the true population of known exoplanets that underline demographics and population synthesis efforts. Finally, vetting serves as a mean to build a knowledge base for upcoming machine-learning codes that will have the necessary function of validating the stream of new candidates.

A number of pipelines have been developed to validate photometric signals caused by transiting exoplanets, such as \robovetter \citep{Coughlin2020} and \autovetter \citep{Jenkins2014}. These codes are based on supervised machine learning -- a single decision tree for the former and a random forest algorithm for the latter -- and have been trained to recognize true planet candidates based on tens of thousands of human-inspected signals available at the time of development. Other algorithms, like \vespa \citep{Morton2015} and \triceratops \citep{Giacalone2021}, compute the Bayesian probability that the signal might be a false positive based on the number of nearby sources in TESS field of view, their properties, as well as the shape and strength of the investigated signal. Yet another approach, pioneered by \astronet \citep{Shallue2018} for Kepler and exploited by \astroramjet \citep{Olmschenk2021} for TESS, utilizes a deep learning algorithm known as a Convolutional Neural Network (CNN). The code is trained to recognize patterns in light curves and discriminate among classes based on the data alone. Finally, \cite{Kostov2019} used the Discovery And Vetting of Exoplanets (\dave) pipeline that tests the candidate signal at both the pixel and light curve levels mimicking the process that expert human vetters apply. \citet{Kostov2019} made use of \dave to uniformly vet 776 K2 candidates and collected the data and dispositions in a catalog hosted on the Kepler Threshold Crossing Event Review Team (TCERT) \href{http://keplertcert.seti.org/DAVE/}{website}. Since then, \dave has been adapted to TESS data and used to vet a handful of individual planet candidates, see e.g. \citet{Kostov2019_toi175}, \citet{Gilbert2020}. 

Here we present the TESS Triple-9 (TT9) catalog of 999 uniformly-vetted exoplanet candidates detected in TESS data. The candidates were identified by the community using the SPOC \citep{Jenkins2016} and QLP \citep{huang2020} pipelines, and are listed on ExoFOP-TESS. To create the catalog, we vet the signals employing the \dave pipeline and citizen science. Specifically, each candidate is analysed through \dave and inspected by human vetters, and also utilizing auxiliary information as provided by ExoFOP TESS\footnote{https://exofop.ipac.caltech.edu/tess/}. This effort has been possible also thanks to the Planet Patrol project (Kostov et al., submitted), a NASA-led citizen science project hosted on Zooniverse through which a group of citizen scientists interested in validating TESS planets joined our science team. The main goal of Planet Patrol is to evaluate the reliability of \dave results with the help of volunteers. The outcomes of the classifications have been implemented in \dave to rule out bad images and compute high-fidelity photocenter statistics. Soon after the launch of the project, several citizen scientists expressed interest in further helping the vetting efforts and, after joining the science team, quickly became proficient vetters. These "Superusers" have been key to the success of this endeavor. The Superusers and the science team members will be referred hereafter as "vetters". 

This paper is organized as follows. We present our vetting workflow in Section 2; the details on the citizen science project Planet Patrol are described in Section 3; the TESS Triple-9 catalog is presented in Section 4. Finally, we wrap up the conclusions in Section 5. 

\begin{figure*}
    \centering
    \includegraphics[width=0.9\linewidth]{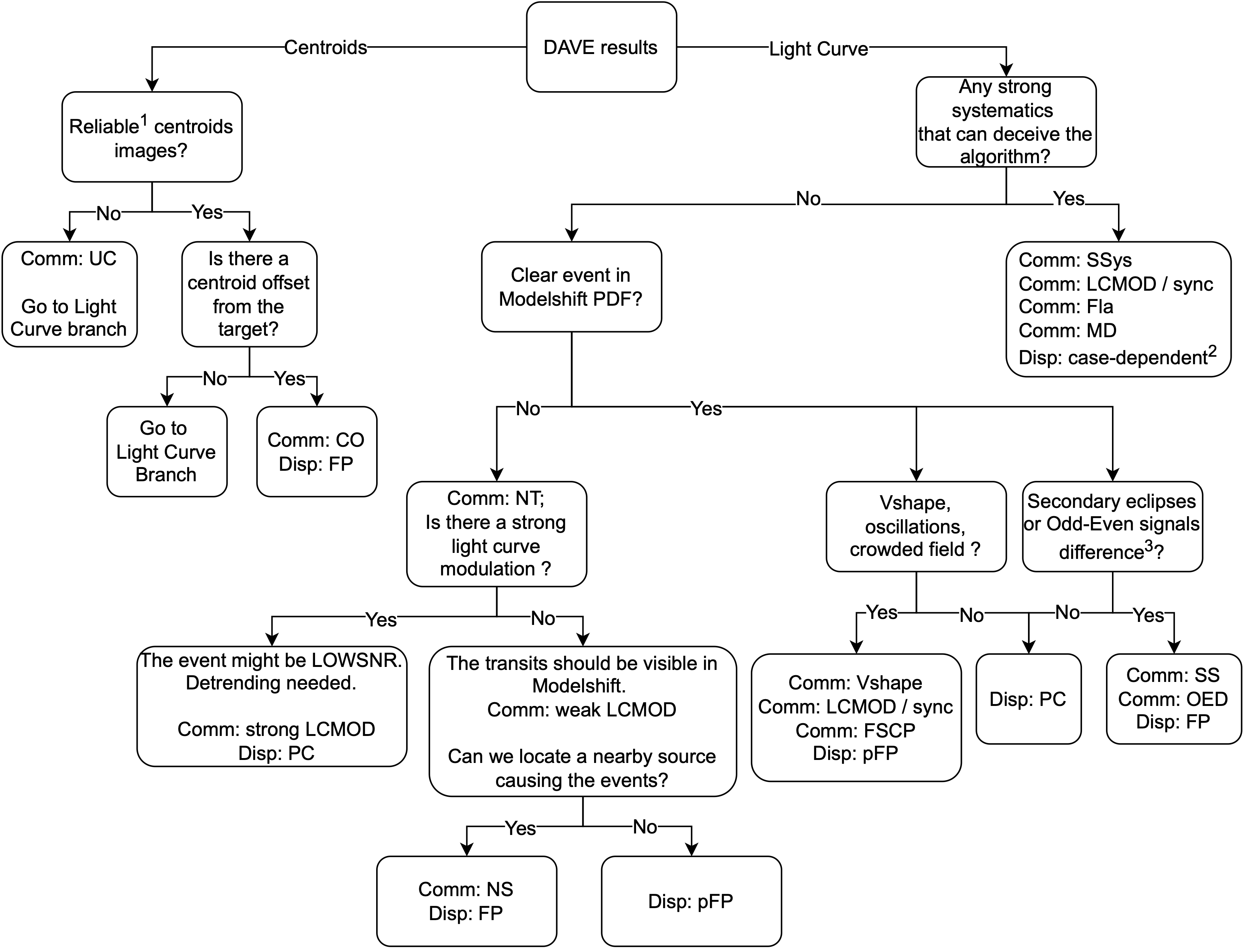}
    \caption{A flow diagram showing the algorithmic process the vetters would go through during the vetting of the TOIs. The workflow starts with the left "Centroids" branch and moves on to the "Light Curve" branch if needed, as explained in Section \ref{workflow}. All of the used abbreviations are shown in Table \ref{tab:abbreviations}.\\ 
    Footnote$^1$: centroids are considered reliable if no other field bright variable source is present. Footnote$^2$: the mentioned features could make it difficult to interpret the nature of the signal without jeopardizing its true planetary nature \textit{a priori}. Footnote$^3$: Given that these secondary eclipses and/or odd-even differences are not caused by any irregular feature in the light curve (previously inspected).}
    \label{fig:diagram}
\end{figure*}

\section{The vetting workflow}
\label{workflow}

For every TOI listed in our TT9 catalog, we performed uniform vetting analysis following a workflow that is based on the interpretation of \dave results using as input the transit ephemeris, duration, and depth as provided by ExoFOP-TESS. Where available, we also evaluate ancillary information that the pipeline does not account for such as archival data. The workflow proceeds as outlined in the next subsections, it is summarized in Fig.\ref{fig:diagram} and it is mainly based on products generated by the \dave pipeline that we describe below.

\begin{itemize}
    \item \textsf{Photocenter}: this module of \dave generates the in-transit and the two (before and after the event) out-of-transit images for each transit. Then, it subtracts the in-transit image from the average out-of-transit image to produce a difference image. Finally, it measures the center-of-light for each difference image by fitting the TESS pixel-response-function (PRF) and a Gaussian point-spread-function (PSF) to the image. This is highlighted in Fig.\ref{fig:centroidimage} for the case of TIC 43647325, where a fainter field star is present a few pixels above and to the right of the target (upper right panel). This field star is, however, not present in the difference image, demonstrating that it is not varying in brightness during the detected transit events. The measured photocenters are on-target (black star; TIC 43647325), confirming that it is the source of the detected transits.
    The overall photocenter (red circle) is calculated by averaging over the individual photocenters (red dots) corresponding to each detected transit as examined in the previous step (see Fig.\ref{fig:centroidimage}).
    See \citet{Kostov2019} for further details.
    \item \textsf{Modelshift}: this module of DAVE phase-folds the light curve and convolves it with a trapezoid model using the transit parameters provided on ExoFOP-TESS, thus highlighting light curve features that resemble the detected transits but occur at orbital phases other than zero. The module is designed to highlight the average of the input primary signals, the average of the odd and even signals, the most prominent secondary, tertiary and positive features. The results of the module are summarized in an automatically-generated PDF file (see Fig. \ref{fig:modshiftex}). See \citet{Kostov2019} for further details.
    \item \textsf{Lomb-Scargle}: \dave runs a Lomb-Scargle (LS) periodogram \citep{Lomb1976}, \citep{Scargle1982} and generates a PDF file showing both the light curve phase-folded on the period of the inspected signal and the best-fit LS period (see Fig.\ref{fig:why_check_lc}). This test is performed to help the vetter check whether the modulations occur on the same (or half) period of the candidate signal. In such cases, the vetter might comment for potential beaming, ellipsoidal, and/or reflections effects that might indicate a binary star system, see \citep{Morris1993}, \citep{faigler2011}, \citep{Shporer2017}. Performing this test is the reason we opt out of applying custom detrending of the light curves. 
    \item \textsf{Summary}: \dave produces a summary PDF file that contains the full light curve, a zoom-in of each transit in the light curve, the \textsf{photocenter} module images and the LS result.
\end{itemize}
Besides \dave results, the vetters consulted stellar catalogs such as SIMBAD \citep{Wenger2000} and Gaia EDR3 \citep{gaia2021} to keep track of the field stars in TESS images. This is critical to assess whether known nearby field stars could potentially be the true source of the observed TESS signals.
Accompanying \dave products and our final dispositions, we kept track of any noticeable feature for each TOI by means of comments. The details of these are listed in Table \ref{tab:abbreviations}.

\begin{table*}
\centering
\begin{tabular}{p{0.10\linewidth} | p{0.20\linewidth} | p{0.60\linewidth}}
\hline
Abbreviation & Meaning & Description \\ \hline \hline
Disposition & &      \\ \hline 
PC & Planetary Candidate & A TOI that passed all vetting tests. \\

(p)FP & False Positive & A TOI that does not (fully) pass the vetting tests. \\ \hline \hline
Comments &  &       \\ \hline
(p)SS & Significant Secondary & A statistically significant secondary is highlighted by the \textsf{Modelshift} module. A secondary eclipse is typical of an eclipsing binary star.  \\ \\

(p)CO & Centroid Offset & A significant photocenter shift is detected. This indicates that the target star is not the source of the investigated signal. \\ \\

UC & Unreliable Centroids & A centroid image is considered unreliable if stray light, bright field stars or low SNR cause the difference image to be too noisy for proper photocenter measurements. \\ \\

(p)OED & Odd-Even Difference & A statistically significant difference between odd and even eclipses is highlighted by the \textsf{Modelshift} module and/or the vetters, indicating an eclipsing binary star.\\ \\ 

(p)Vshape & V-shape & The shape of the signal is not U-shaped, as expected from a typical transit, but rather V-shaped. A transiting planet (usually with $R << R_{star}$) is expected to produce a sharp ingress, a flat bottom, and a sharp egress. In contrast, an eclipsing star (usually with $R \sim R_{star}$) often produces gradual ingress and egress. \\ \\

TD & Too Deep & For a given stellar radius, the deeper the eclipse, the larger the eclipser. As the deepest confirmed exoplanet eclipses range between 3-4\%\footnote[1]{\ref{note1}}, e.g. \citep{Noyes2008},\citep{Triaud2013},
we flag signals with depth greater than 2.5\%. We note that we do not use TD as the only indicator for a (p)FP as the stellar radius provided in by ExoFOP-TESS might be under/overestimated. \\ \\

FSCP & Field Star in Central Pixel & TESS pixels cover a sky-projected area of 21"x21" each. This means that other sources might fall within the same pixel of the target and be the true source of the signal. \\ \\ 

LCMOD & Light Curve MODulation & Oscillations in the light curve due to intrinsic and/or rotational variability (potentially due to pulsations or spots) that are not synchronized with the orbital period. These can be produced by either the target itself of by a nearby field star that falls in the aperture used to extract the light curve. Such lightcurves are generally not indicative of a potential false positive. \\ \\

sync & Synchronous Modulations & Lightcurve variations synchronized with the TOI orbital period (or half) are common features of variable binary stars. For example, stars in binary systems with small orbital separations can be deformed into ellipsoids due to tidal effects. The light curves of these systems can show sinusoidal oscillations on half of the eclipses period due to changes in the light-emitting area of the components along the orbital phase, \citep{Shporer2017}. Such lightcurves are generally indicative of a potential false positive. \\ \\

EB & Eclipsing Binary & An eclipsing binary system. \\ \\

HPMS & High Proper Motion Star & A high proper motion star as listed on SIMBAD\footnote{http://simbad.u-strasbg.fr/simbad/}.\\ \\

SSys & Strong Systematics & The long-cadence \textsf{eleanor} light curve displays strong artifacts.\\ \\

Fla & Flares & TESS light curve shows potential flaring events. These are sharp rises in flux followed by an exponential decay. \\ \\

NT & No Transit & The \textsf{eleanor} light curve does not show noticeable transit-like signals for QLP-detected TOIs. The \textsf{Modelshift} module flags the candidate due to the low statistical significance of the expected signal. \\ \\

NS & Nearby Source & We identify a nearby source that is the true source of the signal. This has been achieved using the \textsf{eleanor pixel\_by\_pixel()} function, as explained in Section 4.2. \\ \\

MD & Momentum Dump & TESS thruster-firing-induced artifact in the light curve. \\ \\

LOWSNR & Low Signal to Noise Ratio & The SNR of the expected transits is too low for reliable vetting.\\ \hline
\end{tabular}
\caption{List of abbreviations used during the vetting process. The first part of the table shows the final disposition abbreviation. The second half displays the general comments that have been used to support the disposition. A \textsf{(p)} is used next to certain comments and means \textsf{potential}. The human vetters would add a \textsf{(p)} in case \dave did not automatically flag the feature due to lack of statistical significance.}
\label{tab:abbreviations}
\end{table*}

\subsection{Photocenter Analysis}

\begin{figure}
    \centering
    \includegraphics[width=\linewidth]{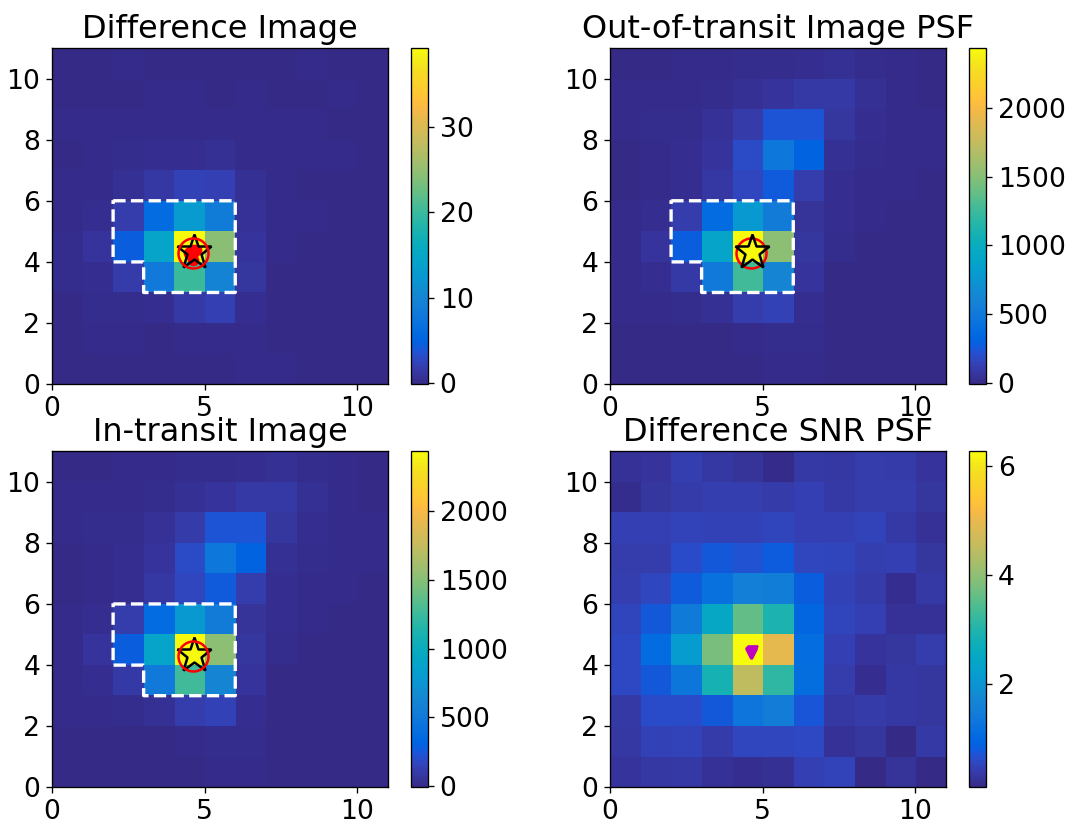}
    \caption{\dave photocenter analysis for TIC 43647325. Upper right: the average out-of-transit TESS image. The white contour represents the pixels used to extract the light curve, the black star represents the catalog position of the target, the purple triangle represents the measured average out-of-transit photocenter, and the red circle represents the measured difference image photocenter; lower left: the average in-transit image; upper left: difference image highlighting the location of the source (black star symbol) varying in brightness during the detected transits.
    The colorbars units are in electrons per second ($e^-/s$).
    Taking into account \dave results as a whole, this TOI has been marked as a planet candidate (PC).} 
    \label{fig:centroidimage} 
\end{figure}

As a first step, the vetter would start with the critical problem of pinpointing the true source of the transit-like events. Individual TESS pixels cover a sky-projected area of ${\rm 21 \times 21}$ arcseconds$^2$ . This often means that one or more field stars fall in the same pixel as the target or in the aperture used to extract the light curve. First, the vetter would inspect stellar charts and catalogs to evaluate whether any known sources fall within the aperture. Next, they would check if these sources are bright enough to cause the transit-like signal in the light curve based on the measured transit depth and on the magnitude difference between the target and the source. For every target, we considered a threshold $\Delta mag$ such that only field stars with a magnitude within this limit could cause a signal with the same depth of the one under consideration. 
In cases for which one or more such sources are identified, the final disposition for the TOI is accompanied by the comment FSCP (Field Star in Central Pixels). This procedure is complementary to the photocenter inspection, during which the vetter would examine the TESS difference images generated by \dave and identify the true source of the signal (see Fig.\ref{fig:centroidimage}).

If the difference images present a photocenter shift away from the target, the candidate signal is considered a false positive since the signal is coming from a different star and the disposition is accompanied by the comment 'CO' (Centroid Offset). If the corresponding difference images show a well-defined group of pixels centered on the target, the candidate signal would pass the centroid analysis. There are cases for which the centroid difference images are difficult to interpret due to artifacts, stray-light or low SNR. When this was observed to be the case, vetters would comment with 'UC' (Unreliable Centroids). 

We note that small exoplanets could produce low SNR signal in TESS data which would also result in poor photocenter images. If the light curves did not not show any other red flag, we would flag the poor quality of the measured photocenters but still label the signal as planetary candidates (PC). \\

\subsection{Light Curve Analysis}
\begin{figure*}
    \centering
    \includegraphics[width=\linewidth]{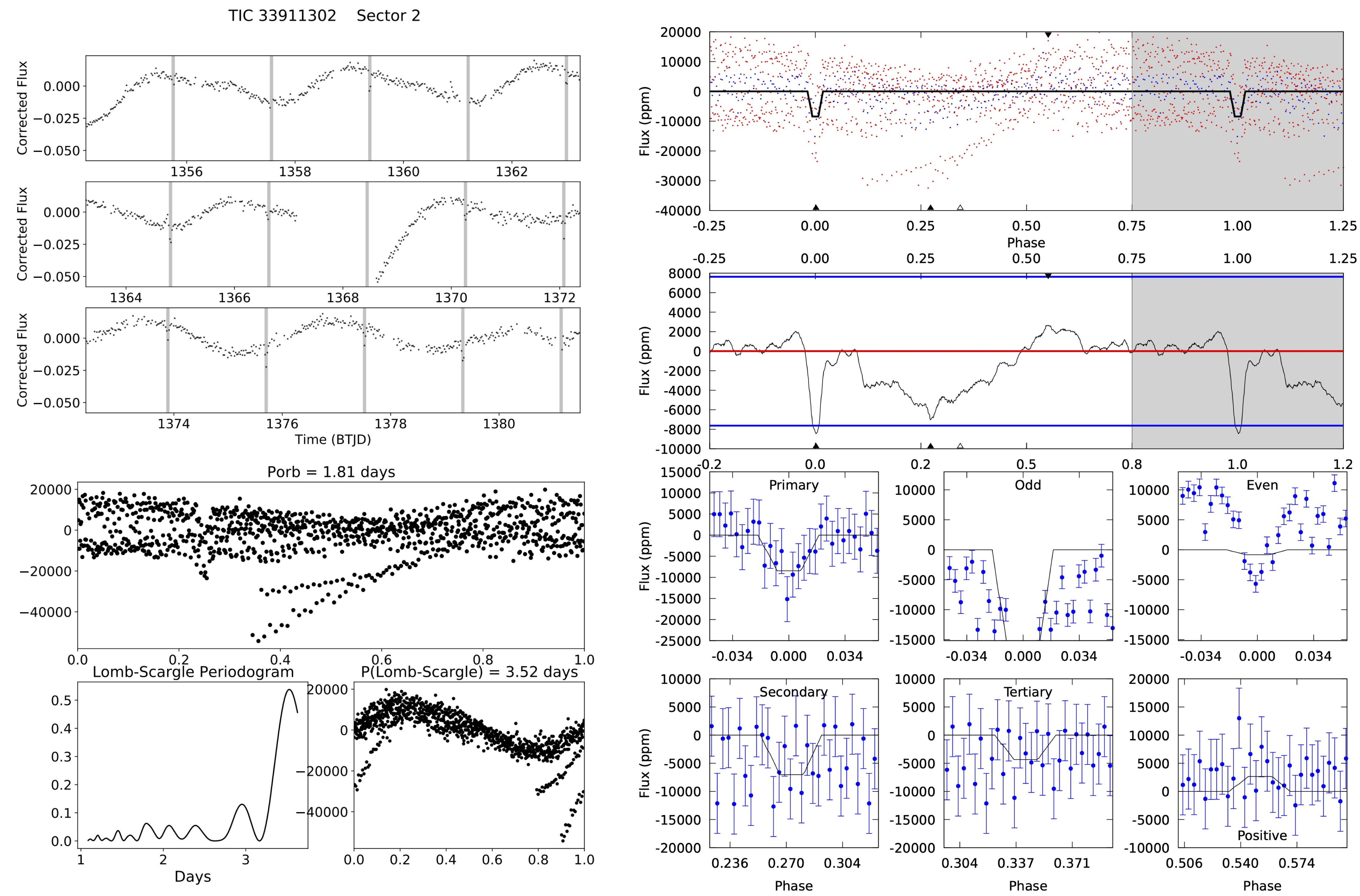}
    \caption{Prominent stellar variability seen in the light curve of TIC 33911302 (upper left panel). The variability is quantified by \dave through a Lomb-Scargle analysis and provided to the vetter as a phase-folded light curve (lower left panels). Such light curve variability can result in nominal false positive disposition by the \textsf{Modelshift} module (right panel) due to measured odd-even difference -- but can be overruled by the vetter as an incorrect automated disposition. The first panel of the \textsf{Modelshift} shows the phase-folded lightcurve along with the best-fitting transit model (black line); the second panel of the \textsf{Modelshift} plot shows the same as the upper panel but the light curve is convolved with the transit model; the lower panels shows zoom-ins on the primary and secondary events (as labeled), the odd and even primary events, as well as any tertiary or positive events. The uppermost table displays the significance of the aforementioned features, see \citet{Kostov2019} for the details.
    This TOI has been labelled as a PC because the variability is not reminiscent of beaming, reflection and/or ellipsoidal effects. The observed oscillations could be due to stellar variability of the target itself or of another source that falls within the group of pixels used to extract the light curve. There is no hint of a secondary eclipse standing out of the light curve noise. We note that since the stellar variability for this target is not coherent with the orbital period, the ``Vshape'' flag alone is not sufficient to indicate a TOI as a potential false positive.}
    \label{fig:why_check_lc}
\end{figure*}

Once the measured photocenters have been analysed, we inspect the full light curve for every available sector. It is worth noting that we use \eleanor ``corrected flux'' light curves \citep{feinstein2019} without any processing such as custom detrending or removal of any data point.
This visual inspection is necessary to assess whether prominent systematics could make \dave automated dispositions unreliable. During our work, unreliable detections were mostly observed when using TESS Full Frame Images-extracted light curves. Since \dave uses un-detrended data by design, this step is also critical in situations where strong stellar variability is present. The variability can throw off the evaluation process of some features such as the depth difference between odd and even signals (see Fig.\ref{fig:why_check_lc}). The light curve variability is also quantified with a Lomb-Scargle periodogram and a phase-folded light curve is provided as part of a summary PDF for every TIC (as in Fig.\ref{fig:why_check_lc}). 

Thus, the full light curve and the LS PDFs aid the vetters in evaluating the \textsf{Modelshift} result while taking into account potential irregular behavior in the light curve. 

This step of the analysis results in comments about the light curve and the signals added to the final disposition. Typical comments used in this phase are LCMOD, Fla, MD, WE and SSys (see Tab.\ref{tab:abbreviations} for the details).

The next step is to inspect the Modelshift PDF, in which the folded light curve and the primary and second-order features are highlighted.

The vetter would inspect the \dave evaluations on top of the PDF (see Fig.\ref{fig:modshiftex}), listed in a table where numbers in red indicate potential issues for the respective candidate signal (see \citet{Kostov2019} for details.). This table could highlight significant additional eclipses, odd-even depth difference or even positive spikes. However, vetters have been trained to evaluate the issues raised by \textsf{Modelshift}  based on what they have seen in the previous step, i.e. the full light curve inspection. For example, when the light curves present a modulation, the vetters do not automatically interpret an odd-even difference flagged by \dave as a strong indicator of a false positive. Instead, the vetters would point out the necessity to detrend the light curve to obtain a final disposition based on the consistency of the depth of consecutive signals. Again, during the \textsf{Modelshift} inspection, vetters could identify features with the comments listed in Table \ref{tab:abbreviations}.


\subsection{Final Dispositions and Comments}
For each target, at least three vetters, plus one science team member, have performed a complete and thorough vetting as described, in an effort to minimize the subjectivity of the process by mediating among different dispositions. The results of this effort are described in Section \ref{catalog}. 
The final dispositions presented in our catalog are as follows:

\begin{itemize}
    \item a planet candidate (PC) is a TOI that passes all of the described vetting tests.
    \item a false positive (FP) is a TOI that does not pass one or more tests. It is one with unambiguous non-planetary nature. 
    \item a potential false positive (pFP) is a TOI for which multiple issues have been identified by \dave, by the vetters, or by both -- yet we cannot securely rule it out as a a clear false positive because, for example, the signal-to-noise ratio is low, or the signal is coming from a pixel containing comparably-bright multiple sources. 
\end{itemize}

To clarify, an example of a pFP might be a TOI presenting a potential secondary eclipse in an otherwise flat light curve (pSS, defined as an additional eclipse that does not stands out beyond the blue $1\sigma$ noise level lines in the \textsf{Modelshift} PDF, see Fig. \ref{fig:pfp_1}. A second example is a candidate signal that exhibits prominent V-shaped eclipses and light curve modulations coherent with the orbital period of the TOI (see Fig. \ref{fig:pfp_2}) yet no significant secondary eclipses or centroid offset were found. Furthermore, a candidate for which there are no discernible transit-like features in the lightcurve is also flagged as a pFP. In contrast to pFPs, an FP disposition is representative of a clear secondary eclipse on the same period of the primary, odd-even transit depth differences, or a photo-center shift during the detected transits. Along with these dispositions, we also provide comments on any other noticeable features using the abbreviations listed in Tab. \ref{tab:abbreviations}.

\section{Citizen Scientists-led Development of Additional Vetting Tools and Resources}
\label{patrol}

The Planet Patrol citizens scientists have not only been fundamental to the vetting process, but have also been responsible for major contributions to the presentation and visualization of the results. 

LC led a team of volunteers to produce an introductory \href{https://www.youtube.com/watch?v=zt3FeqMar9M&list=PL89HT0OfBC7roVxTQ4GdsQbOUdm9JxoDl&index=12}{Video Tutorial} hosted on YouTube, in order to help lower the bar for further involvement in the vetting process by the wider community. The video introduces the key concepts about the search for transiting exoplanets and describes the vetting process through examples.

HADL has produced video recordings of all of our meetings, which can serve as further introduction to vetting as well as informal tutorials, recommendations, guidance, resources, etc. for newcomers. The recordings are posted on \href{https://www.youtube.com/playlist?list=PL9W9aPp3UQUxCPofULvfPb4bcUcFucG0C}{Youtube}. If there is a sufficient interest in the community, we will migrate these recordings to a dedicated repository.

Yet another project that emerged from the superuser group (led by RS, a student at Staples High School) is the development of a custom web-based interface designed to streamline the vetting process -- in essence, a dedicated vetting portal. Currently, we are using Google Sheets to collect the dispositions from each vetter for each target, which involves manual input of thousands of entries (e.g. using the abbreviations listed in Table 1 plus free text). This is a slow and sometimes cumbersome process prone to errors. To address this issue, we are transitioning to a custom Graphical User Interface which allows user-friendly drop-down menus, multiple-choice answers, free text, automated uploading of targets that still need to be vetted, etc. For completeness, we briefly describe the development of the vetting portal below.

The website was written in Typescript React, is hosted by Heroku at \href{https://planet-patrol-web.herokuapp.com/paper}{Planet Patrol Website}, and stores data in an IBM Cloud; the source code can be found at \href{https://github.com/Orion31Dev/Planet-Patrol-Vetting-Website}{GitHub}. While designing the website, it was important to be mindful of the workflow that vetters were accustomed to. For example, the vetting portal has to A) present information in a condensed and easy-to-navigate layout; B) allow users to find relevant information regarding each target, including the corresponding ExoFOP-TESS link and the PDF files produced by \dave in PDF format; C) let users write dispositions on each target, giving them as much freedom in their comments while requiring a machine-readable format; D) facilitate effective analysis of user dispositions. The portal expanded the user experience by including options to easily sort the table by parameters, find targets by number of dispositions, and by providing direct links to the vetting PDFs produced by \dave (using the Google Drive API to automatically search for the relevant document). Also, the portal offers drop-down menus to input dispositions, buttons with the pre-defined comments of Table \ref{tab:abbreviations} and free-text input -- all of which can be accesses in machine-readable format. Lastly, the new website includes features that automatically highlight targets that still need to be vetted by the user, the dispositions and comments of other users, as well as the final group dispositions. 

\begin{table*}
    \centering
    \resizebox{\linewidth}{!}{%
    \begin{tabular}{|l|l|l|l|l|l|l|l|l|l|l|l|l|l|}
    \hline
        TIC ID & Sectors & Epoch [BTJD] & Period [d] & Duration [hrs] & Depth [ppm] & $R_p$ [$R_J$] & $R_s$ [$R_\odot$] & $TESS_{mag}$ & Disposition & Comment\\ \hline \hline
        1003831 & 8 & 2458518.203 & 1.651142 & 0.76 & 3007 & 0.49 & 0.98 & 10.6701 & PC & TOI 564 b \\ \hline
        1103432 & 8 & 2458519.87 & 3.727891 & 3.87 & 17864 & 1.66 & 1.34 & 12.8283 & pFP & pTD, pVshape, FSCP, pSS \\ \hline
        1129033 & 4 & 2458410.985 & 1.360025 & 2.16 & 16381 & 1.09 & 0.95 & 9.62713 & PC & WASP-77 A b \\ \hline
        1133072 & 8 & 2458517.532 & 0.846542 & 2.41 & 2525 & 0.15 & 0.33 & 12.625 & pFP & FSCP , LCMOD, short-P \\ \hline
        1449640 & 5 & 2458440.434 & 3.50175 & 3.80 & 14123 & 1.98 & 1.70 & 11.921 & FP & SS \\ \hline
        1528696 & 5 & 2458438.406 & 0.88202 & 0.58 & 9750 & 1.50 & 0.78 & 13.1686 & PC & NGTS-6 b \\ \hline
        2758565 & 2 & 2458356.031 & 3.78192 & 2.04 & 15390 & 2.35 & 1.17 & 12.3611 & pFP & pSS,TD \\ \hline
        4616072 & 6 & 2458469.062 & 4.18599 & 2.85 & 12942 & 1.70 & 1.56 & 12.8893 & PC & HATS-45 b \\ \hline
        4646810 & 4 & 2458416.346 & 14.490034 & 1.59 & 905 & 0.23 & 0.78 & 8.8723 & PC & ~ \\ \hline
        4897275 & 21 & 2458872.341 & 16.710042 & 5.40 & 615 & 0.24 & 1.09 & 7.6474 & PC & HPMS \\ \hline
        5109298 & 7 & 2458491.797 & 1.6221 & 2.43 & 2310 & 0.94 & 2.22 & 10.6698 & FP & Vshape, CO \\ \hline
        5772442 & 7 & 2458492.502 & 1.10578 & 1.87 & 673 & 0.67 & 2.62 & 10.351 & FP & NS \\ \hline
    \end{tabular}%
}
    \caption{The final result of our work is a table of dispositions for 999 TESS Objects of Interest (TOIs). Here, we report an extract of it. For each TOI, we display the TESS Input Catalog (TIC) identifier of the star, the signal parameters input in \dave, the radius of the potential transiter, the radius of the star, the TESS magnitude of the star and the final disposition with related comments. The full table is available as an electronic supplement.}
    \label{tab:final_table}
    \end{table*}

\subsection{Lessons Learned}
As part of this work, we kept record of aspects that could be useful for similar citizen science efforts. For the benefit of the community, here we share our experience. \\

1) Zooniverse talk board is a convenient method for engaging with the volunteers. We tried to respond to every question as quickly as possible and also provided a Google Form for them to exploit their interest (if any) and ask to join the group of SuperUsers.\\

2) Only about a third of the volunteers who signed up for further involvement in the project were active and participated in the weekly meetings. We suspect part of the reason is the time difference, as outlined below. \\

3) Regular interactions between the science team and the SuperUsers were vital for the success of the effort. For offline discussions, we used a dedicated Google Group and a Slack channel; both worked well. For live discussions, we used Google Meet. However, scheduling a weekly meeting that worked for people spread across the world was difficult and we believe this would be an issue for any citizen science project that attract the interest of people worldwide. While the North/South American and European/African time zones could be covered simultaneously from a single location, volunteers from the Asian/Australian time zones were practically left out due to the time difference. One solution to this issue could be to have two separate weekly meetings, led from two locations separated by 12 hours. \\

4) Keeping a video record of the meetings provided a convenient catch-up option for SuperUsers who could not participate in the meetings. \\

5) An easy to use, intuitive, and comprehensive tutorial is crucial to get volunteers who join further along during the project up to speed in a reasonable time frame. The resources and tools developed as part of the Planet Patrol project, and the information gathered and synthesized, could be used in the future as such tutorials. \\

6) The information provided by ExoFOP-TESS, Gaia, MAST, Simbad, Vizier, and other publicly-available databases was an important contribution for the success of the project. It allowed volunteers to not only gain deeper understanding of the particular target they were vetting but, as a natural consequence, also of general astronomical concepts such as spectrosopic binary stars, parallax, proper motion, SED, etc.

\section{The TESS Triple-9 Catalog}
\label{catalog}
The TT9 catalog contains half the TOIs listed on ExoFOP TESS as of October 2020; we plan to present the rest of our vetting analysis in future work. The TOIs have been detected by the \textsf{spoc} and \qlp pipelines using the 2-minute and 30-minute cadence data. The specifics of these pipelines are presented in \citet{Jenkins2016} and \citet{huang2020}. Our analysis led to the vetting of 709 signals as true planetary candidates (PC), i.e. TOIs that have passed all our vetting tests. 
We identified 144 TOIs as FPs and 146 as potential false positives (pFP). The dispositions have been collected in a table with period of the eclipses, epoch, depth, duration, stellar and TOI estimated radii, the stellar TESS magnitude and general comments (see Table \ref{tab:final_table}).

\begin{figure*}
    \centering
    \includegraphics[width=0.9\linewidth]{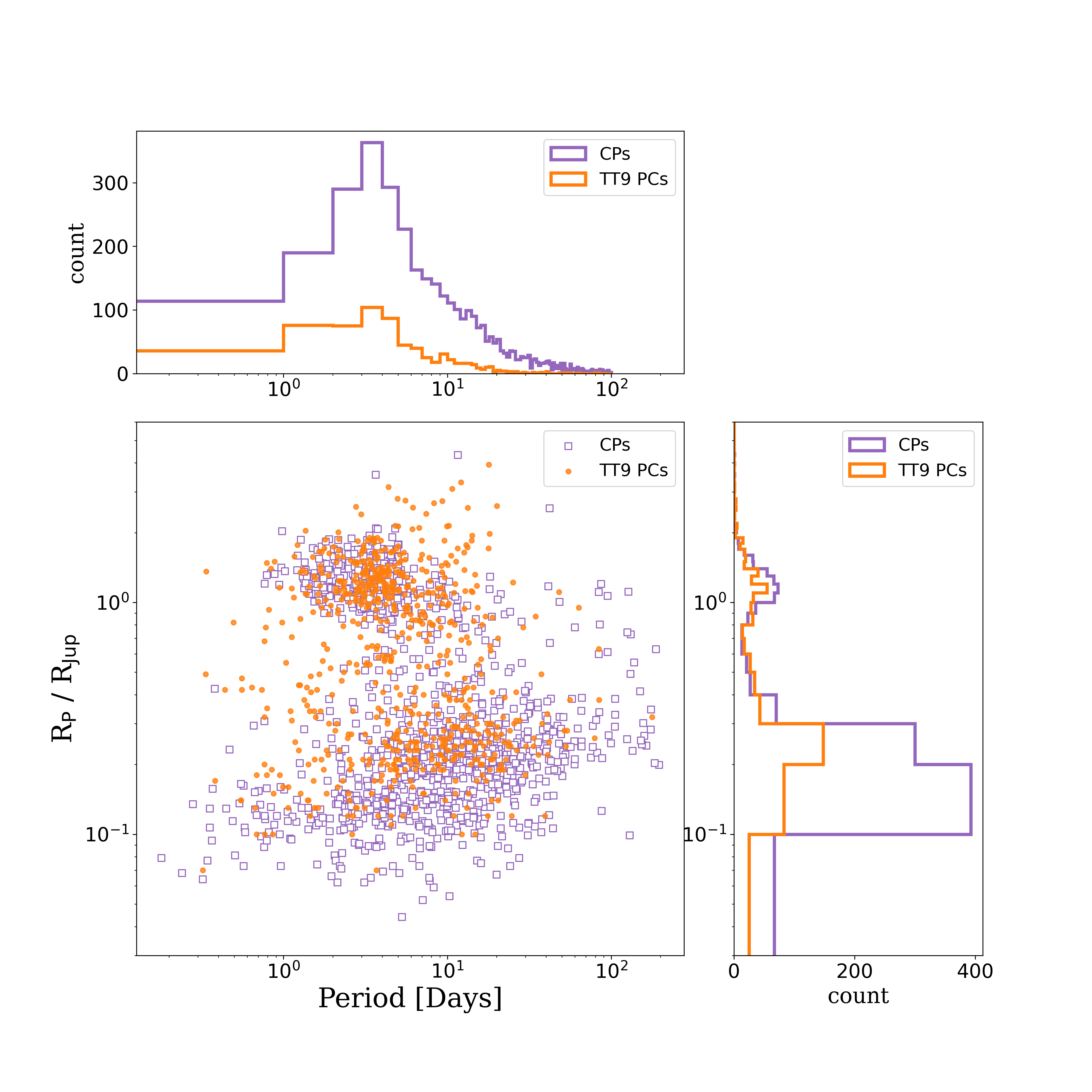}
    \caption{Comparison between the 709 planetary candidates in the TT9 catalog (orange circles) with a subset of the entirety of confirmed exoplanets (purple squares). We report confirmed exoplanets discovered via the precise radial velocities and transits techniques with a period lower than 200 days for clarity.}
    \label{fig:PCvsCP}
\end{figure*}

\subsection{Planetary Candidates}

Of the 709 TOIs marked as PCs in TT9, 146 have already been confirmed according to ExoFOP TESS. The largest among these are hot Jupiters discovered by ground-based projects like the Wide Angle Search for Planets (WASP) \citep{Pollacco2006}, the Hungarian-made Automated Telescope Network project (HAT-N) \citep{Bakos2013}, the XO effort \citep{McCullough2005}, the Qatar Exoplanet Survey (QAS) \citep{Alsubai2013}, the Kilodegree Extremely Little Telescope survey \citep{Pepper2004} and the Next Generation Transit Survey \citep{Wheatley2018}. The smallest planets are instead discoveries of space-based missions: CoRoT \citep{Catala1995}, Kepler \citep{Borucki2008}, K2 \citep{Howard2015} and TESS. 

The most common comments for our PCs are ``Field Star in Central Pixel'' (FSCP, 81 times), ``Light Curve MODulation'' (LCMOD, 77 times) and ``V-shaped signal'' (Vshape, 55 times). The first is expected since TESS cameras pixels cover a ${\rm 21 \times 21}$ arcseconds area on the sky, resulting in a frequent blend of multiple sources. The second can either be caused by modulation of the target star itself or from sources that fall within the  aperture used to extract the light curve. Finally, a V-shaped transit is not a conclusive proof against the planetary nature of a transiter, but it suggests that the dimensions of the two celestial objects at play are comparable. This is expected to happen for binary star systems more than a planet-star pair due to geometrical reasons even though large hot Jupiters with grazing transits have also been shown to produce this type of signal, e.g. \citep{Smalley2011}, \citep{Mancini2014}, \citep{Bento2017}.  We note that we label as PC also 4 TOIs for which we see no obvious signals in the light curves displaying strong modulations (commented with ``NT, strong LCMOD''). In these cases the modulations are strong enough to hide the weak transit signal without detrending, which we do not perform \textsf{ad hoc}. Hence, we keep these as PCs since a signal might have been actually detected if the data were properly detrended.

The final PC yield of the TT9 catalog is shown in Fig.\ref{fig:PCvsCP}, in terms of orbital period and planet radius, and also compared to a subset of confirmed exoplanets. We only display confirmed exoplanets that have been discovered via the radial velocity and transit photometry techniques, with periods shorter than 200 days for clarity (3,884 out of the total 4,935 confirmed exoplanets listed on the NASA Exoplanet Archive, as of mid-January 2022). 
A depression in the number of planetary candidates is observed for the TT9 PCs as well as for the confirmed exoplanets when {$0.3 R_J < R < 1.0 R_J$}.

\begin{figure*}
    \includegraphics[width=\linewidth]{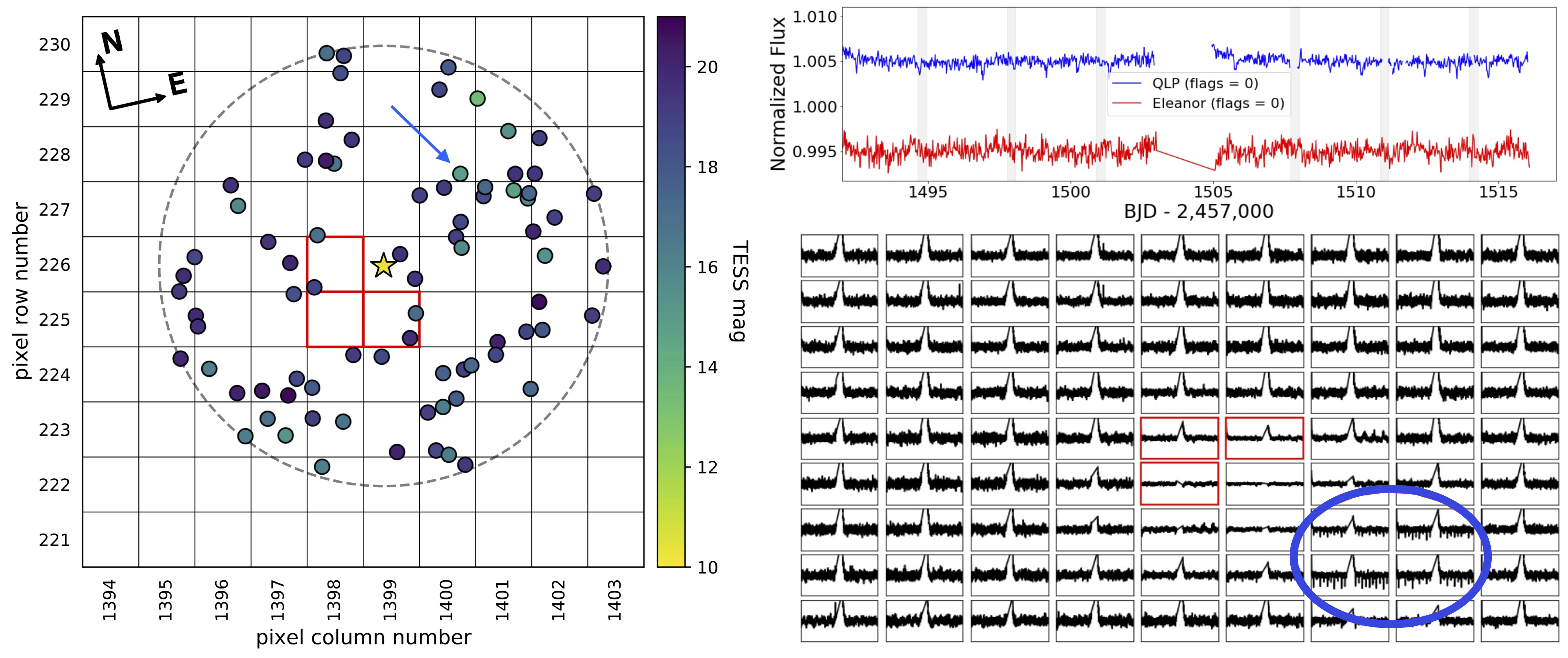}
    \caption{The left panel shows the 10x10 TESS pixels centered on the target (yellow star), the field stars brighter than $T_{mag}$ = 21 mag (dots) within a 4-pixel radius, the \textsf{eleanor} aperture (red solid outline), and the source of the signals detected in QLP (blue dashed arrow). The colorbar show the TESS magnitude of the field sources. The image has been created using the code \textsf{triceratops} \citep{Giacalone2021}.
    Upper right panel: TESS sector 7 QLP (blu) and eleanor (red) light curves for TIC 5772442. While the transits of the TOI are clear with a period of approximately 2.5 days in the QLP light curve, no sign of them is present in the \textsf{eleanor} data. In this case, we use the \textsf{eleanor} pixel-by-pixel light curves (lower right panel) to find that the signal is coming from a nearby source (blue outline in upper right panel) while the light curve extracted for the target using the eleanor aperture (red solid outline in middle panel) is flat.  This TOI is thus a false positive (FP). We note that (i) the \textsf{eleanor} aperture does not include the target star; and (ii) the lower panel is flipped along the horizontal axis compared to the lower right panel.}
    \label{fig:qlp_vs_eleanor}
\end{figure*}

\subsection{False Positives}

A total of 144 TOIs in our catalog have been flagged as false positives (FPs). The most common reason is a significant photocenter shift during transit -- 76 targets -- while secondary eclipses have been registered for 48 candidate signals. In 6 cases the poor quality of the light curve might have deceived the detection pipeline that locked onto features like gaps in the sectors, momentum dump spikes or other systematics (TICs 24364065, 47384844, 101929303, 150247134, 169177766, 198384408). Only 2 TOIs show significant odd-even differences between consecutive transits (TIC 9033144 and TIC 230086768). Light curve modulations are present in 32 cases. Of these, 8 display modulations coherent with the orbital period (TICs 2758565, 93963408, 96246348, 97158538, 141663460, 141663464, 198457103, 233720539). We kept track of such modulations for the following reasons. First, in some cases they could be triggering the detection pipelines and be mistaken for candidate planets. Secondly, light curve modulations coherent with the orbital period of the transiting exoplanet candidate can be indicative of a short-period eclipsing binary. Finally, even though the detected modulation is not the deciding factor for the FP disposition, we include it for the sake of completeness.

We note that in some cases more than one FP red flags were raised.

Another frequent source of false positives we identified is non-detections of QLP-based planet candidates in \eleanor light curves. As mentioned above, our analysis is based on the latter because vetting transiting planet candidates detected by one pipeline but using light curves produced by another is highly valuable as it provides an independent set of tests and and confirmations (e.g. Kostov et al. 2019)\footnote{We note that the QLP light curves were not publicly-available at the start of this project. As we aim for consistency and uniformity instead of completeness, we continued our analysis with \eleanor data even after the QLP light curves were uploaded to MAST.} Specifically, \dave uses \eleanor-generated ``Corrected Flux'' light curves to perform its tests on the TOIs for which only FFI data is available. For some targets in our catalog, the \eleanor light curves do not show obvious transit signals at the ephemeris provided by ExoFOP-TESS. In these cases, the vetters would label them as being potential false positives rather than certain false positives. This is mainly because the apparent lack of transit signals in \eleanor data can be caused by differences between the QLP pipeline (through which the signals where discovered) and the \eleanor pipeline (through which we vet them). Altogether, there are 53 such ``No Transit'' (NT) cases. We have already mentioned why 4 of them have been labeled as PC in the previous subsection.

We note that because there is no apparent transit-like signal for these NT TOIs in \eleanor data, we cannot perform a photocenter analysis. Thus to analyse them and produce a final disposition, we visually inspected the corresponding light curves using the \eleanor \textsf{pixel\_by\_pixel()} routine. This function allows the user to inspect the light curves of single pixels within the downloaded TESS pixel cutout. Specifically, we would compare the \eleanor aperture used to extract the light curve to the single pixels light curves. The fact that {\tt eleanor} uses a different aperture optimization algorithm implies that the final light curve is extracted from a different set of pixels and might be different with respect to the QLP one. 

Indeed, we found 14 (out of the 53 NT cases) for which the QLP-detected signal is coming from a nearby source (NS) instead of the target under investigation and we finally label these as false positives. We report an example of this case in Fig. \ref{fig:qlp_vs_eleanor}. 

Finally, we note that 6 confirmed planets have been labelled in TT9 as FPs due to the presence of significant secondary eclipses -- despite the fact that these are confirmed hot Jupiters. These dispositions, however, are based on our definition of an FP, i.e. a candidate exhibiting clear secondary eclipses (see Fig.\ref{fig:occ}). Thus even though we are aware that these `secondary eclipses' are in fact planetary occultations, for consistency we flag the 6 TOIs as FPs (see also Table \ref{tab:occultation} and Section 4.4).

\subsection{Potential False Positives}
\begin{figure}
    \centering
    \includegraphics[width=\linewidth]{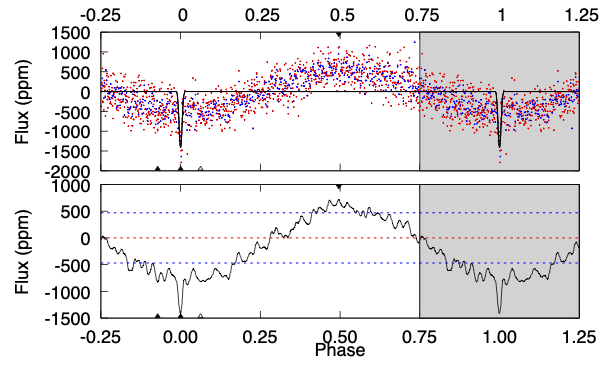} \
    \includegraphics[width=\linewidth]{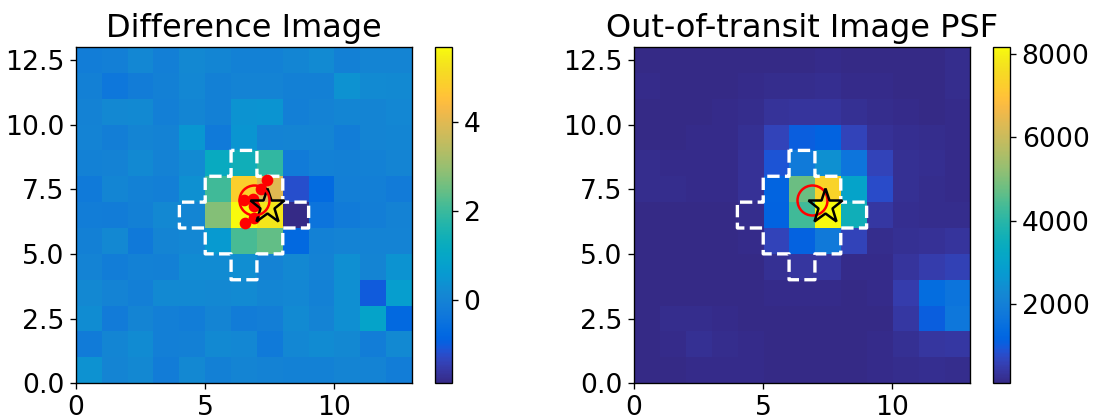}
        \caption{An example of potential false positive (pFP): TIC 97158538. The upper panel presents a portion of the \textsf{Modelshift} results. Here, one can see that the TOI presents light curve modulations coherent with eclipses. The lower panels displays the out-of-transit TESS image (lower right) and the difference image (lower left) between in- and out-of-transit frames. The colorbars show units of electrons per second ($e^-/s$). Given the large scatter in the measured photocenters, likely caused by the light curve variability, we note a potential, sub-pixel centroids offset.}
    \label{fig:pfp_2}
\end{figure}

As described in Section \ref{workflow}, we have labelled as potential false positives (pFP) 146 TOIs for which we could not confirm the suspected non-planetary nature yet identified a number of potential issues. For these TOIs, we identified various combinations of: V-shape morphology, ultra-short periods, potential secondary eclipses, potential photocenter offset, field stars in the same pixel of the target that are bight enough to produce the detected signals as contamination, and/or coherent light curve modulations as commonly seen in short-period eclipsing binaries. We reported examples of potential false positives that could be due to unresolved background sources in Fig.\ref{fig:pfp_2}  and Fig.\ref{fig:pfp_1}. Programs such as TFOP\footnote{TESS Follow-Up Observing Programs} could clear these cases thanks to higher-resolution photometric measurements, spectroscopic observations, high-resolution imaging or precise radial velocity measurements. Specifically, we identified 49 cases showing potential photocenter offset and 32 cases for which a potential secondary eclipse has been identified. A total of 69 pFPs display light curve modulations (LCMOD), of which 33 are synchronous with the eclipses (sync). Finally, 35 TOIs show no transit event but, as opposed to the FP cases, we could not identify nearby sources showing the expected eclipses. 

In some cases, more than one of these red flags have been raised by the vetters.

\begin{table}
    \centering
    \begin{tabular}{c|c|c|c|}
         TIC & Name & $\sim$Depth (ppm) & Comments\\ \hline
         16740101 & KELT9-b & 600 & Occ\\
         22529346 & WASP-121 b & 200 & Occ\\
         86396382 & WASP-12 b & 400 & FSCP, Occ\\
         100100827 & WASP-18 b & 300 & Occ \\
         129979528 & WASP-33 b & 500 & LCMOD, Occ \\
         158324245 & KOI-13 b & 400 &   Occ\\
    \end{tabular}
    \caption{Confirmed planets in the TT9 catalog for which our vetting efforts identify prominent secondary eclipses. According to our vetting workflow, these planets had to be flagged as false positives in TT9  because we do not have sufficient information to distinguish between an occultation and a secondary eclipse.}
    \label{tab:occultation}
\end{table}

\subsection{Confirmed Planets}
\label{CPs}
For consistency, we vet targets independent of any prior knowledge of the system, and regardless of their current disposition on ExoFOP. Some of them are listed as confirmed planets and were added to our catalog for completeness. Our analysis of confirmed planets supports the planetary nature interpretation of 140 of them and flagged 6 as false positives. The reason for the latter dispositions is as follows.

When large, close-in planets pass behind their host star along the line of sight with the observer, a secondary dip in the light curve can be observed due to the occultation of the light reflected and/or emitted by the planet. Occultations allow direct measurement of the planet's radiation (hence its effective temperature) and helps to constrain the planet's orbital eccentricity. Our analysis highlighted this kind of signal as a false positive indicator since the event is effectively identical to a secondary eclipse. We find such occultations in the light curves of 6 of the confirmed planets in the TT9 and we label these signals as false positives since we vetted every TOI without prior knowledge of its current status. An example is shown in Fig. \ref{fig:occ}. We report the TICs of these candidates, along with the approximate depth of the secondary eclipse as seen in TESS bandpass (600-1000 nm), in Table \ref{tab:occultation}. For completeness, we also point out the other disposition comments for these targets where applicable.

\subsection{Potential New Signals}

During the analysis presented above, the vetters picked up new signals not previously reported on ExoFOP TESS. We inspect these one by one and check whether they could be identified with known systematic effects such as Momentum Dumps (MD), i.e. artifacts caused by TESS thruster firings. This can be done by comparing the signals to the registered MDs events in TESS Data Releases\footnote{\url{https://archive.stsci.edu/tess/tess\_drn.html}}. These new signals are as follows. 

\begin{figure}
    \centering
    \includegraphics[width=\linewidth]{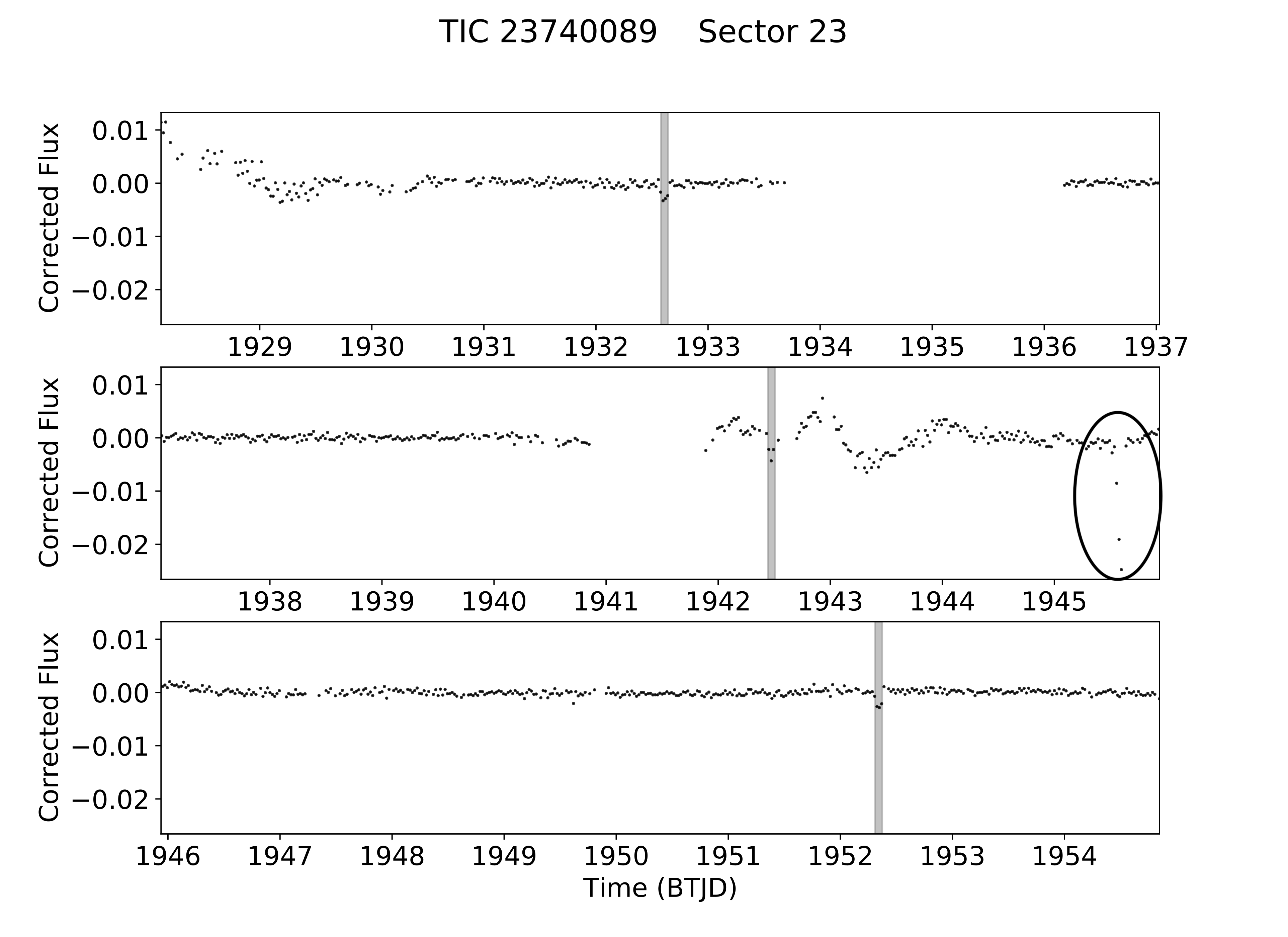}
    \caption{Long-cadence light curve of TIC 23740089. The highlighted signals are the events detected by the QLP pipeline that we vet with \dave. An additional eclipse is observed at BTJD 1945.5. If this were the primary eclipse of a circular binary system, such that the detected TOI is in fact secondary eclipses, we would have to observe the same type of feature at approximately 1935.6 and 1955.3. Neither time was observed due to a data gap and the finite time range of the TESS sector so we cannot conform this hypothesis.}
    \label{fig:extra1}
\end{figure}

An additional eclipse has been flagged for TIC 23740089, located at BTJD 1945.5 in sector 23. This TIC hosts what we have labelled as a PC with a 0.59 $R_J$ candidate orbiting a 1.14 $R_\odot$ star in 9.8 days. The additional, deeper ($>2\%$) eclipse is only observed once between transits two and three (see Fig. \ref{fig:extra1}). According to TESS Data Release Notes for sector 23, this event does not correspond to a thruster firing event. The separation between the TOI and this additional signal is such that we could be missing the same deeper eclipse event after transit number one due to a data gap and the one after transit number three due to the finite duration of the sector. Thus, the available \textsf{eleanor}-extracted TESS data does not have sufficient coverage to assess whether this event might indicate a binary star and we labeled the target as a PC with an extra eclipse. TESS will observe again the target between February 26th 2022 and April 22nd 2022 which might allow to clear this specific case. 
\begin{figure}
    \centering
    \includegraphics[width=\linewidth]{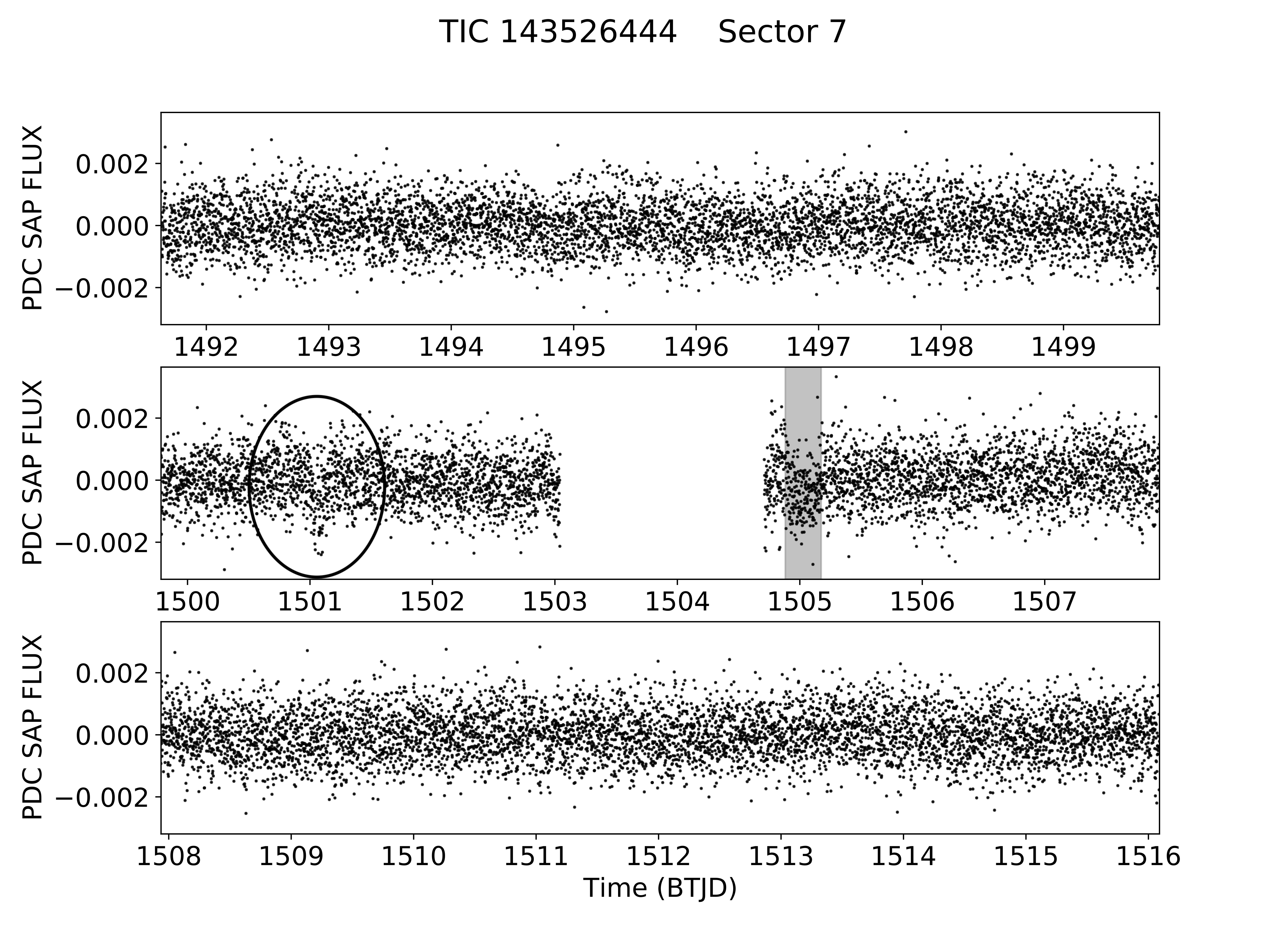}
    \caption{Short-cadence sector 7 light curve of TIC 143526444. The vertical gray band highlights the transit of the candidate in the ExoFOP TOIs catalog. The additional event at BTJD 1501 (black oval) is likely caused by a momentum dump as reported in TESS data release notes. We don't find any similar event in the other available sectors.}
    \label{fig:extra3}
\end{figure}

We detected two additional events in the light curve of TIC 74534430 at approximately 1539.6 and 1545.0 BTJD (in sector 8 and 9) that do not fall on any reported MD events. These events are shallower and shorter than the TOI transits, which is labelled in TT9 as a PC with an orbital period of 18 days and radius of 0.19 $R_J$, orbiting a 0.61 $R_\odot$ star. The target has been observed by TESS in sectors 8, 9, 35 and 36. 

We also observed additional, non-periodic events in the light curve of TIC 128501004 in sector 16, the only one available for the target. No additional TESS observations are scheduled for this star. Two different deep events have been noticed in the light curve at approximately 1747.9 and 1748.2 (BTJD), neither of which corresponds to a MD event. TIC 128501004 is a 1.1 $R_\odot$ star hosting a 0.18 $R_J$ object (labelled as a PC in TT9) with an orbital period of 0.8 days. However, taking into account the additional events, this candidate could be a triple star system which will require further observations to be confirmed. 

An additional event is present in the light curve of TIC 143526444 in sector 7. The target is a 3.5 $R_\odot$ star hosting a 0.9 $R_J$ PC with an orbital period of 15.3 days. The event is at time 1501 (BTJD) where a MD event has been registered and reported in TESS Data Release Notes $\#9$. Thus we reject this extra event as an artifact.

Finally, a potential new candidate has been observed in the TESS sector 4 light curve of TIC 100990000 at an epoch of approximately 1425 (BTJD). This transit-like feature is a single event within the four available TESS sectors (3, 4, 30 and 31) and has a duration of roughly six hours which is consistent with a long period. If confirmed, this candidate would be the outermost planet in a system with other two planets (Earth- and Super-Earth-sized) orbiting a G-type star in 4.04 and 9.57 days, respectively. The estimated radius of the new candidate is 2.7 $R_\oplus$. The characterization of this system is being carried out in Cacciapuoti et al, submitted. We note that we have recently learned that this candidate has also been uploaded on ExoFOP-TESS as TIC 100990000.03 by a citizen scientist of the Planet Hunters TESS project \citep{Eisner2021}.

\section{Conclusions}

We presented the TT9 catalog, containing 999 uniformly-vetted transiting exoplanet candidates from TESS. We marked 709 TOIs as \textsf{bona fide} planet candidates, of which 146 are confirmed exoplanets. Additionally, 144 TOIs were flagged as false positives due to photocenter motion during transit, significant secondary eclipses, and/or light curve modulations indicative of eclipsing binary stars. Finally, 146 TOIs were labelled as potential false positives due to too many potential issues to be passed as PCs yet no definitive evidence to be ruled out as clear false positives.

The TT9 catalog is provided to the community in a table format (see Table \ref{tab:final_table}) which provides our final dispositions along with additional comments for each TOI. For completeness, the table also includes the transit ephemeris, depth, and duration, and the TESS magnitude. 
The \dave generated \textsf{modelshifts} and PDF summaries, as well as the photocentre images, are made publicly available on ExoFOP TESS at the dedicated webpage of each TOI.
This work presents the first stage of our effort to vet all TOIs listed on ExoFOP-TESS using \dave and citizen science. Our results provide an independent analysis of known TOIs and could be used to prioritize follow-up observations. The dispositions provided in the TT9 catalog can be also utilized as input for demographic studies of transiting exoplanets from TESS or as a training set for machine learning algorithms aimed at automating vetting efforts.

\section{Acknowledgements}

This research has made use of the NASA Exoplanet Archive, which is operated by the California Institute of Technology, under contract with the National Aeronautics and Space Administration under the Exoplanet Exploration Program.

We acknowledge the use of public TESS Alert data from the pipelines at the TESS Science Office and at the TESS Science Processing Operations Center (SPOC) and from the Massachusetts Institute of Technology Quick Look Pipeline (QLP). 

This research has made use of the Exoplanet Follow-up Observation Program website, which is operated by the California Institute of Technology, under contract with the National Aeronautics and Space Administration under the Exoplanet Exploration Program.

This publication uses data generated via the Zooniverse.org platform, development of which is funded by generous support, including a Global Impact Award from Google, and by a grant from the Alfred P. Sloan Foundation.

We thank the referee for the insightful comments which helped improve the manuscript.

\textit{Software}: \dave \citep{Kostov2019}, \eleanor \citep{feinstein2019}, \qlp \citep{huang2020}

\section{Data Availability}
The data underlying this article will be shared on reasonable request to the corresponding author.

\bibliographystyle{mnras}
\bibliography{biblio}

\begin{thebibliography}{}
\makeatletter
\relax
\def\mn@urlcharsother{\let\do\@makeother \do\$\do\&\do\#\do\^\do\_\do\%\do\~}
\def\mn@doi{\begingroup\mn@urlcharsother \@ifnextchar [ {\mn@doi@}
  {\mn@doi@[]}}
\def\mn@doi@[#1]#2{\def\@tempa{#1}\ifx\@tempa\@empty \href
  {http://dx.doi.org/#2} {doi:#2}\else \href {http://dx.doi.org/#2} {#1}\fi
  \endgroup}
\def\mn@eprint#1#2{\mn@eprint@#1:#2::\@nil}
\def\mn@eprint@arXiv#1{\href {http://arxiv.org/abs/#1} {{\tt arXiv:#1}}}
\def\mn@eprint@dblp#1{\href {http://dblp.uni-trier.de/rec/bibtex/#1.xml}
  {dblp:#1}}
\def\mn@eprint@#1:#2:#3:#4\@nil{\def\@tempa {#1}\def\@tempb {#2}\def\@tempc
  {#3}\ifx \@tempc \@empty \let \@tempc \@tempb \let \@tempb \@tempa \fi \ifx
  \@tempb \@empty \def\@tempb {arXiv}\fi \@ifundefined
  {mn@eprint@\@tempb}{\@tempb:\@tempc}{\expandafter \expandafter \csname
  mn@eprint@\@tempb\endcsname \expandafter{\@tempc}}}

\bibitem[\protect\citeauthoryear{{Alsubai} et~al.,}{{Alsubai}
  et~al.}{2013}]{Alsubai2013}
{Alsubai} K.~A.,  et~al., 2013, \actaa, \href
  {https://ui.adsabs.harvard.edu/abs/2013AcA....63..465A} {63, 465}

\bibitem[\protect\citeauthoryear{{Bakos} et~al.,}{{Bakos}
  et~al.}{2013}]{Bakos2013}
{Bakos} G.~{\'A}.,  et~al., 2013, \mn@doi [\pasp] {10.1086/669529}, \href
  {https://ui.adsabs.harvard.edu/abs/2013PASP..125..154B} {125, 154}

\bibitem[\protect\citeauthoryear{{Baranne} et~al.,}{{Baranne}
  et~al.}{1996}]{Baranne1996}
{Baranne} A.,  et~al., 1996, \aaps, \href
  {https://ui.adsabs.harvard.edu/abs/1996A&AS..119..373B} {119, 373}

\bibitem[\protect\citeauthoryear{Barclay, Pepper  \& Quintana}{Barclay
  et~al.}{2018}]{Barclay_2018}
Barclay T.,  Pepper J.,   Quintana E.~V.,  2018, \mn@doi [The Astrophysical
  Journal Supplement Series] {10.3847/1538-4365/aae3e9}, 239, 2

\bibitem[\protect\citeauthoryear{{Bento} et~al.,}{{Bento}
  et~al.}{2017}]{Bento2017}
{Bento} J.,  et~al., 2017, \mn@doi [\mnras] {10.1093/mnras/stx500}, \href
  {https://ui.adsabs.harvard.edu/abs/2017MNRAS.468..835B} {468, 835}

\bibitem[\protect\citeauthoryear{{Borucki} et~al.,}{{Borucki}
  et~al.}{2008}]{Borucki2008}
{Borucki} W.,  et~al., 2008, in {Sun} Y.-S.,  {Ferraz-Mello} S.,   {Zhou}
  J.-L.,  eds, ~ Vol. 249, Exoplanets: Detection, Formation and Dynamics. pp
  17--24, \mn@doi{10.1017/S174392130801630X}

\bibitem[\protect\citeauthoryear{{Catala} et~al.,}{{Catala}
  et~al.}{1995}]{Catala1995}
{Catala} C.,  et~al., 1995, in {Ulrich} R.~K.,  {Rhodes} E.~J. J.,   {Dappen}
  W.,  eds,  Astronomical Society of the Pacific Conference Series Vol. 76,
  GONG 1994. Helio- and Astro-Seismology from the Earth and Space. p.~426

\bibitem[\protect\citeauthoryear{{Ciardi}, {Pepper}, {Colon}, {Kane}  \&
  {Astrophysical Community}}{{Ciardi} et~al.}{2018}]{Ciardi2018}
{Ciardi} D.~R.,  {Pepper} J.,  {Colon} K.,  {Kane} S.~R.,   {Astrophysical
  Community} W. I. f.~t.,  2018, arXiv e-prints, \href
  {https://ui.adsabs.harvard.edu/abs/2018arXiv181008689C} {p. arXiv:1810.08689}

\bibitem[\protect\citeauthoryear{{Coughlin}}{{Coughlin}}{2020}]{Coughlin2020}
{Coughlin} J.~L.,  2020, {Robovetter: Automatic vetting of Threshold Crossing
  Events (TCEs)} (\mn@eprint {ascl} {2012.006})

\bibitem[\protect\citeauthoryear{{Eisner} et~al.,}{{Eisner}
  et~al.}{2021}]{Eisner2021}
{Eisner} N.~L.,  et~al., 2021, \mn@doi [\mnras] {10.1093/mnras/staa3739}, \href
  {https://ui.adsabs.harvard.edu/abs/2021MNRAS.501.4669E} {501, 4669}

\bibitem[\protect\citeauthoryear{{Faigler} \& {Mazeh}}{{Faigler} \&
  {Mazeh}}{2011}]{faigler2011}
{Faigler} S.,  {Mazeh} T.,  2011, \mn@doi [\mnras]
  {10.1111/j.1365-2966.2011.19011.x}, \href
  {https://ui.adsabs.harvard.edu/abs/2011MNRAS.415.3921F} {415, 3921}

\bibitem[\protect\citeauthoryear{{Feinstein} et~al.,}{{Feinstein}
  et~al.}{2019}]{feinstein2019}
{Feinstein} A.~D.,  et~al., 2019, {eleanor: Extracted and systematics-corrected
  light curves for TESS-observed stars} (\mn@eprint {ascl} {1904.022})

\bibitem[\protect\citeauthoryear{{Gaia Collaboration} et~al.,}{{Gaia
  Collaboration} et~al.}{2021}]{gaia2021}
{Gaia Collaboration} et~al., 2021, \mn@doi [\aap]
  {10.1051/0004-6361/202039657}, \href
  {https://ui.adsabs.harvard.edu/abs/2021A&A...649A...1G} {649, A1}

\bibitem[\protect\citeauthoryear{{Giacalone} et~al.,}{{Giacalone}
  et~al.}{2021}]{Giacalone2021}
{Giacalone} S.,  et~al., 2021, \mn@doi [\aj] {10.3847/1538-3881/abc6af}, \href
  {https://ui.adsabs.harvard.edu/abs/2021AJ....161...24G} {161, 24}

\bibitem[\protect\citeauthoryear{{Gilbert} et~al.,}{{Gilbert}
  et~al.}{2020}]{Gilbert2020}
{Gilbert} E.~A.,  et~al., 2020, \mn@doi [\aj] {10.3847/1538-3881/aba4b2}, \href
  {https://ui.adsabs.harvard.edu/abs/2020AJ....160..116G} {160, 116}

\bibitem[\protect\citeauthoryear{{Howard}}{{Howard}}{2015}]{Howard2015}
{Howard} A.,  2015, {Discovery and Characterization of Small Planets from the
  K2 Mission}, NASA ADAP Proposal

\bibitem[\protect\citeauthoryear{{Huang} et~al.,}{{Huang}
  et~al.}{2020}]{huang2020}
{Huang} C.~X.,  et~al., 2020, \mn@doi [Research Notes of the American
  Astronomical Society] {10.3847/2515-5172/abca2e}, \href
  {https://ui.adsabs.harvard.edu/abs/2020RNAAS...4..204H} {4, 204}

\bibitem[\protect\citeauthoryear{{Jenkins}, {McCauliff}, {Catanzarite},
  {Twicken}, {Burke}, {Campbell}  \& {Seader}}{{Jenkins}
  et~al.}{2014}]{Jenkins2014}
{Jenkins} J.~M.,  {McCauliff} S.~D.,  {Catanzarite} J.,  {Twicken} J.~D.,
  {Burke} C.~J.,  {Campbell} J.,   {Seader} S.,  2014, in American Astronomical
  Society Meeting Abstracts \#223. p. 206.02

\bibitem[\protect\citeauthoryear{{Jenkins} et~al.,}{{Jenkins}
  et~al.}{2016}]{Jenkins2016}
{Jenkins} J.~M.,  et~al., 2016, in {Chiozzi} G.,  {Guzman} J.~C.,  eds,
  Society of Photo-Optical Instrumentation Engineers (SPIE) Conference Series
  Vol. 9913, Software and Cyberinfrastructure for Astronomy IV. p. 99133E,
  \mn@doi{10.1117/12.2233418}

\bibitem[\protect\citeauthoryear{{Kostov} et~al.,}{{Kostov}
  et~al.}{2019a}]{Kostov2019}
{Kostov} V.~B.,  et~al., 2019a, \mn@doi [\aj] {10.3847/1538-3881/ab0110}, \href
  {https://ui.adsabs.harvard.edu/abs/2019AJ....157..124K} {157, 124}

\bibitem[\protect\citeauthoryear{{Kostov} et~al.,}{{Kostov}
  et~al.}{2019b}]{Kostov2019_toi175}
{Kostov} V.~B.,  et~al., 2019b, \mn@doi [\aj] {10.3847/1538-3881/ab2459}, \href
  {https://ui.adsabs.harvard.edu/abs/2019AJ....158...32K} {158, 32}

\bibitem[\protect\citeauthoryear{{Lomb}}{{Lomb}}{1976}]{Lomb1976}
{Lomb} N.~R.,  1976, \mn@doi [\apss] {10.1007/BF00648343}, \href
  {https://ui.adsabs.harvard.edu/abs/1976Ap&SS..39..447L} {39, 447}

\bibitem[\protect\citeauthoryear{{Lovis} et~al.,}{{Lovis}
  et~al.}{2005}]{Lovis2005}
{Lovis} C.,  et~al., 2005, \mn@doi [\aap] {10.1051/0004-6361:20052864}, \href
  {https://ui.adsabs.harvard.edu/abs/2005A&A...437.1121L} {437, 1121}

\bibitem[\protect\citeauthoryear{{Mancini} et~al.,}{{Mancini}
  et~al.}{2014}]{Mancini2014}
{Mancini} L.,  et~al., 2014, \mn@doi [\aap] {10.1051/0004-6361/201424106},
  \href {https://ui.adsabs.harvard.edu/abs/2014A&A...568A.127M} {568, A127}

\bibitem[\protect\citeauthoryear{{Marcy}, {Butler}, {Vogt}, {Fischer}  \&
  {Lissauer}}{{Marcy} et~al.}{1998}]{Marcy1998}
{Marcy} G.~W.,  {Butler} R.~P.,  {Vogt} S.~S.,  {Fischer} D.,   {Lissauer}
  J.~J.,  1998, \mn@doi [\apjl] {10.1086/311623}, \href
  {https://ui.adsabs.harvard.edu/abs/1998ApJ...505L.147M} {505, L147}

\bibitem[\protect\citeauthoryear{{Mayor} \& {Queloz}}{{Mayor} \&
  {Queloz}}{1995}]{Mayor1995}
{Mayor} M.,  {Queloz} D.,  1995, \mn@doi [\nat] {10.1038/378355a0}, \href
  {https://ui.adsabs.harvard.edu/abs/1995Natur.378..355M} {378, 355}

\bibitem[\protect\citeauthoryear{{McCullough}, {Stys}, {Valenti}, {Fleming},
  {Janes}  \& {Heasley}}{{McCullough} et~al.}{2005}]{McCullough2005}
{McCullough} P.~R.,  {Stys} J.~E.,  {Valenti} J.~A.,  {Fleming} S.~W.,  {Janes}
  K.~A.,   {Heasley} J.~N.,  2005, \mn@doi [\pasp] {10.1086/432024}, \href
  {https://ui.adsabs.harvard.edu/abs/2005PASP..117..783M} {117, 783}

\bibitem[\protect\citeauthoryear{{Morris} \& {Naftilan}}{{Morris} \&
  {Naftilan}}{1993}]{Morris1993}
{Morris} S.~L.,  {Naftilan} S.~A.,  1993, \mn@doi [\apj] {10.1086/173488},
  \href {https://ui.adsabs.harvard.edu/abs/1993ApJ...419..344M} {419, 344}

\bibitem[\protect\citeauthoryear{{Morton}}{{Morton}}{2015}]{Morton2015}
{Morton} T.~D.,  2015, {VESPA: False positive probabilities calculator}
  (\mn@eprint {ascl} {1503.011})

\bibitem[\protect\citeauthoryear{{Naef} et~al.,}{{Naef}
  et~al.}{2010}]{Naef2010}
{Naef} D.,  et~al., 2010, \mn@doi [\aap] {10.1051/0004-6361/200913616}, \href
  {https://ui.adsabs.harvard.edu/abs/2010A&A...523A..15N} {523, A15}

\bibitem[\protect\citeauthoryear{{Noyes} et~al.,}{{Noyes}
  et~al.}{2008}]{Noyes2008}
{Noyes} R.~W.,  et~al., 2008, \mn@doi [\apjl] {10.1086/527358}, \href
  {https://ui.adsabs.harvard.edu/abs/2008ApJ...673L..79N} {673, L79}

\bibitem[\protect\citeauthoryear{{Olmschenk} et~al.,}{{Olmschenk}
  et~al.}{2021}]{Olmschenk2021}
{Olmschenk} G.,  et~al., 2021, \mn@doi [\aj] {10.3847/1538-3881/abf4c6}, \href
  {https://ui.adsabs.harvard.edu/abs/2021AJ....161..273O} {161, 273}

\bibitem[\protect\citeauthoryear{{Pepe} et~al.,}{{Pepe}
  et~al.}{2004}]{Pepe2004}
{Pepe} F.,  et~al., 2004, \mn@doi [\aap] {10.1051/0004-6361:20040389}, \href
  {https://ui.adsabs.harvard.edu/abs/2004A&A...423..385P} {423, 385}

\bibitem[\protect\citeauthoryear{{Pepper}, {Gould}  \& {Depoy}}{{Pepper}
  et~al.}{2004}]{Pepper2004}
{Pepper} J.,  {Gould} A.,   {Depoy} D.~L.,  2004, in {Holt} S.~S.,  {Deming}
  D.,  eds,  American Institute of Physics Conference Series Vol. 713, The
  Search for Other Worlds. pp 185--188 (\mn@eprint {arXiv} {astro-ph/0401220}),
  \mn@doi{10.1063/1.1774522}

\bibitem[\protect\citeauthoryear{{Pollacco} et~al.,}{{Pollacco}
  et~al.}{2006}]{Pollacco2006}
{Pollacco} D.~L.,  et~al., 2006, \mn@doi [\pasp] {10.1086/508556}, \href
  {https://ui.adsabs.harvard.edu/abs/2006PASP..118.1407P} {118, 1407}

\bibitem[\protect\citeauthoryear{{Rauer} et~al.,}{{Rauer}
  et~al.}{2021}]{Rauer2021}
{Rauer} H.,  et~al., 2021, in European Planetary Science Congress. pp
  EPSC2021--90

\bibitem[\protect\citeauthoryear{{Ricker} et~al.,}{{Ricker}
  et~al.}{2015}]{Ricker2015}
{Ricker} G.~R.,  et~al., 2015, \mn@doi [Journal of Astronomical Telescopes,
  Instruments, and Systems] {10.1117/1.JATIS.1.1.014003}, \href
  {https://ui.adsabs.harvard.edu/abs/2015JATIS...1a4003R} {1, 014003}

\bibitem[\protect\citeauthoryear{{Scargle}}{{Scargle}}{1982}]{Scargle1982}
{Scargle} J.~D.,  1982, \mn@doi [\apj] {10.1086/160554}, \href
  {https://ui.adsabs.harvard.edu/abs/1982ApJ...263..835S} {263, 835}

\bibitem[\protect\citeauthoryear{{Shallue} \& {Vanderburg}}{{Shallue} \&
  {Vanderburg}}{2018}]{Shallue2018}
{Shallue} C.~J.,  {Vanderburg} A.,  2018, \mn@doi [\aj]
  {10.3847/1538-3881/aa9e09}, \href
  {https://ui.adsabs.harvard.edu/abs/2018AJ....155...94S} {155, 94}

\bibitem[\protect\citeauthoryear{{Shporer}}{{Shporer}}{2017}]{Shporer2017}
{Shporer} A.,  2017, \mn@doi [\pasp] {10.1088/1538-3873/aa7112}, \href
  {https://ui.adsabs.harvard.edu/abs/2017PASP..129g2001S} {129, 072001}

\bibitem[\protect\citeauthoryear{{Smalley} et~al.,}{{Smalley}
  et~al.}{2011}]{Smalley2011}
{Smalley} B.,  et~al., 2011, \mn@doi [\aap] {10.1051/0004-6361/201015992},
  \href {https://ui.adsabs.harvard.edu/abs/2011A&A...526A.130S} {526, A130}

\bibitem[\protect\citeauthoryear{{Triaud} et~al.,}{{Triaud}
  et~al.}{2013}]{Triaud2013}
{Triaud} A.~H.~M.~J.,  et~al., 2013, \mn@doi [\aap]
  {10.1051/0004-6361/201220900}, \href
  {https://ui.adsabs.harvard.edu/abs/2013A&A...551A..80T} {551, A80}

\bibitem[\protect\citeauthoryear{{Wenger} et~al.,}{{Wenger}
  et~al.}{2000}]{Wenger2000}
{Wenger} M.,  et~al., 2000, \mn@doi [\aaps] {10.1051/aas:2000332}, \href
  {https://ui.adsabs.harvard.edu/abs/2000A&AS..143....9W} {143, 9}

\bibitem[\protect\citeauthoryear{{Wheatley} et~al.,}{{Wheatley}
  et~al.}{2018}]{Wheatley2018}
{Wheatley} P.~J.,  et~al., 2018, \mn@doi [\mnras] {10.1093/mnras/stx2836},
  \href {https://ui.adsabs.harvard.edu/abs/2018MNRAS.475.4476W} {475, 4476}

\makeatother
\end{thebibliography}

\appendix 
\section{\textsf{Modelshift} figures}
\label{sec:appA}
In this Appendix, we display the examples of \dave Modelshift PDF products that have served as examples during the explanation of the general vetting workflow (Section \ref{workflow}) and of the reasons behind the TT9 dispositions (Section \ref{catalog}).

\begin{figure*}
    \centering
    \includegraphics[width=0.8\linewidth]{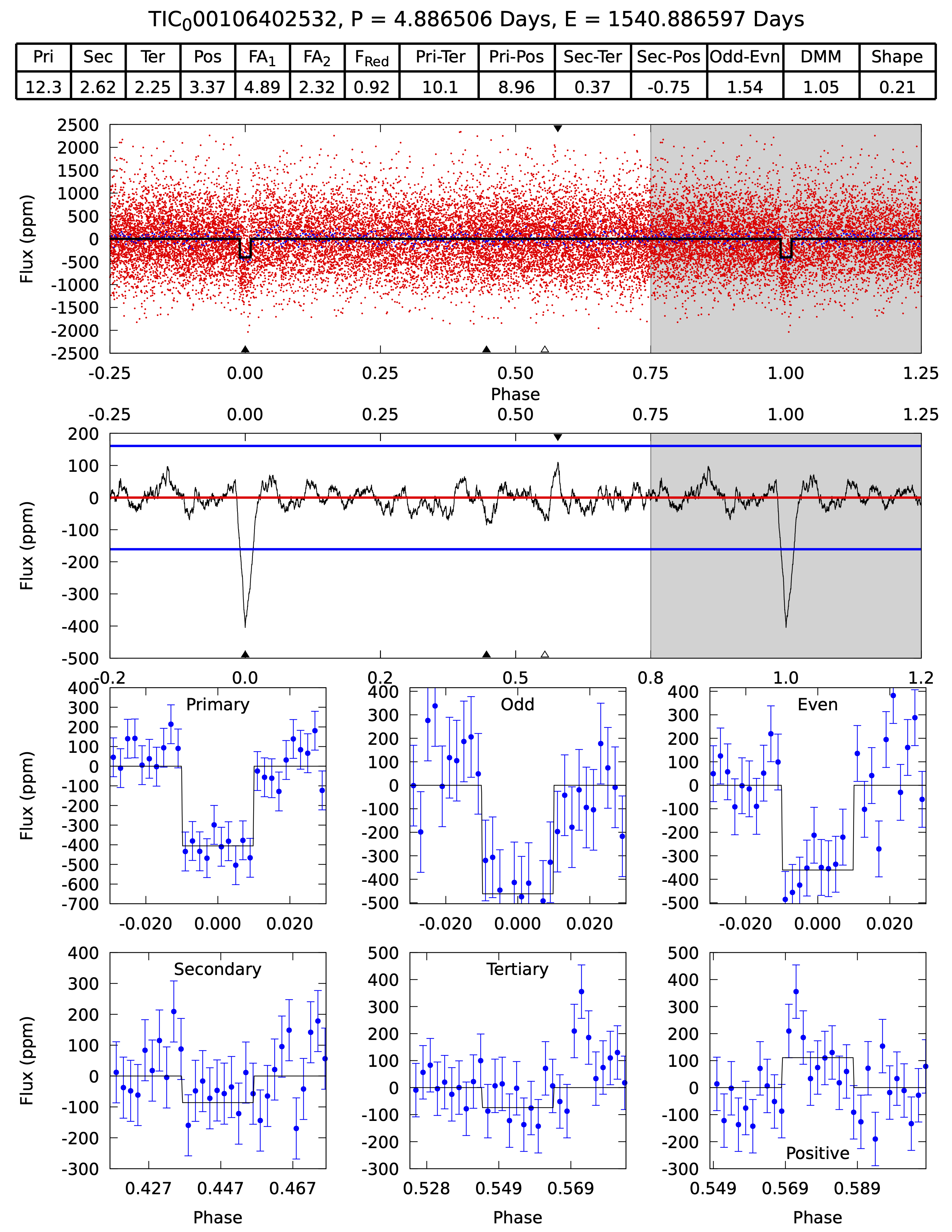}
    \caption{An example of \dave \textsf{Modelshift} PDF for TIC 106402532 (sector 9). No issues of potential concern have been raised by either \dave or the vetters for this TOI. Considering the photocenter analysis as well, this TOI is labelled as a bona-fide planetary candidate (PC) in the TESS Triple-9 catalog.}
    \label{fig:modshiftex}
\end{figure*}

\begin{figure*}
    \includegraphics[width=0.8\linewidth]{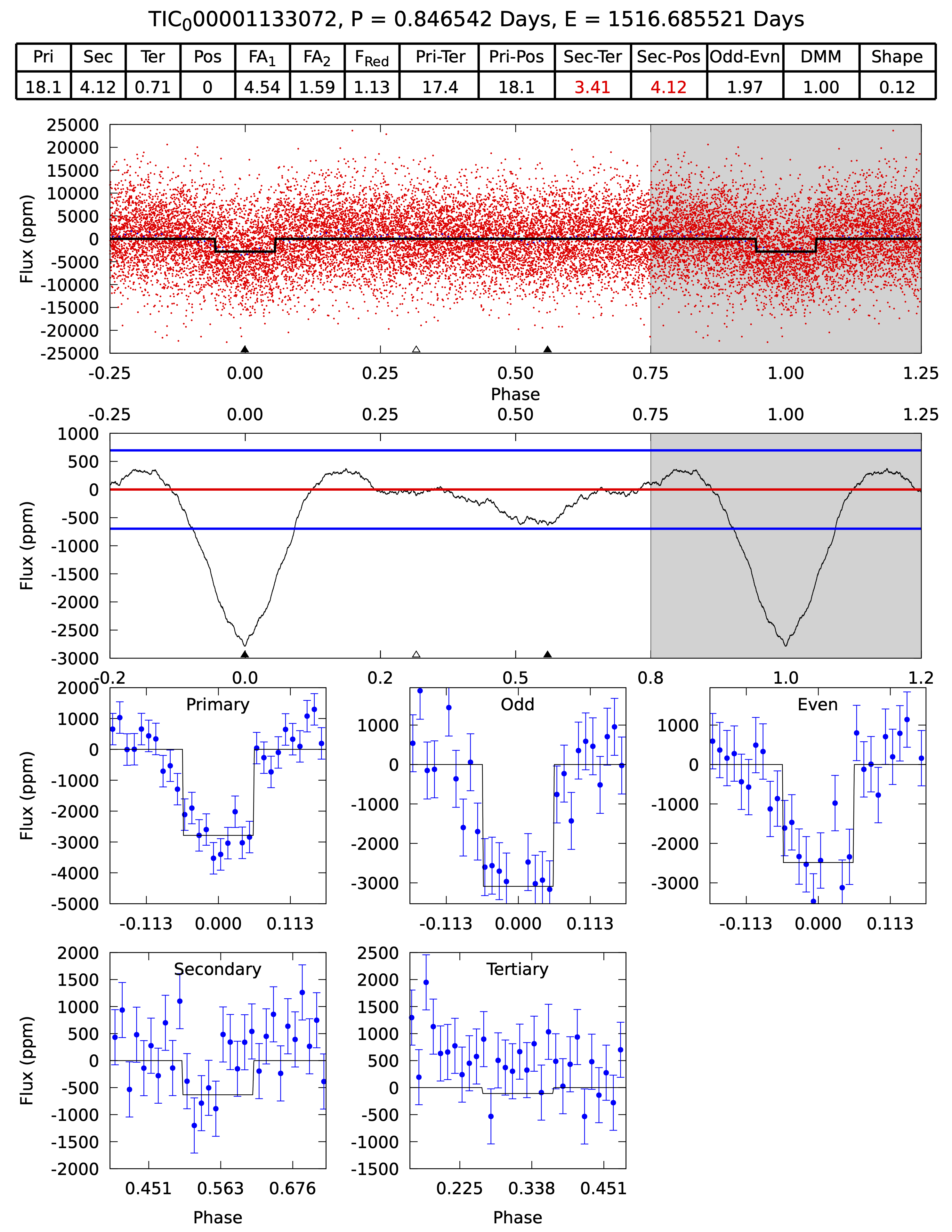} 
    \caption{Modelshift PDF for TIC 1133072 (sector 8). The second panel from top (the phase-folded light curve convolved with the transit model) shows a light curve reminiscent of the modulations typical for $\beta$ Lyrae-type binary stars, where the primary and secondary eclipses do not have sharp ingress and egress. Given the low SNR, the short orbital period, the V-shaped eclipses and the potential secondary eclipse near phase 0.5 (note that it is not flagged as significant by \dave), we flag the target as a pFP.}
    \label{fig:pfp_1}
\end{figure*}

\begin{figure*}
    \centering
    \includegraphics[width=0.8\linewidth]{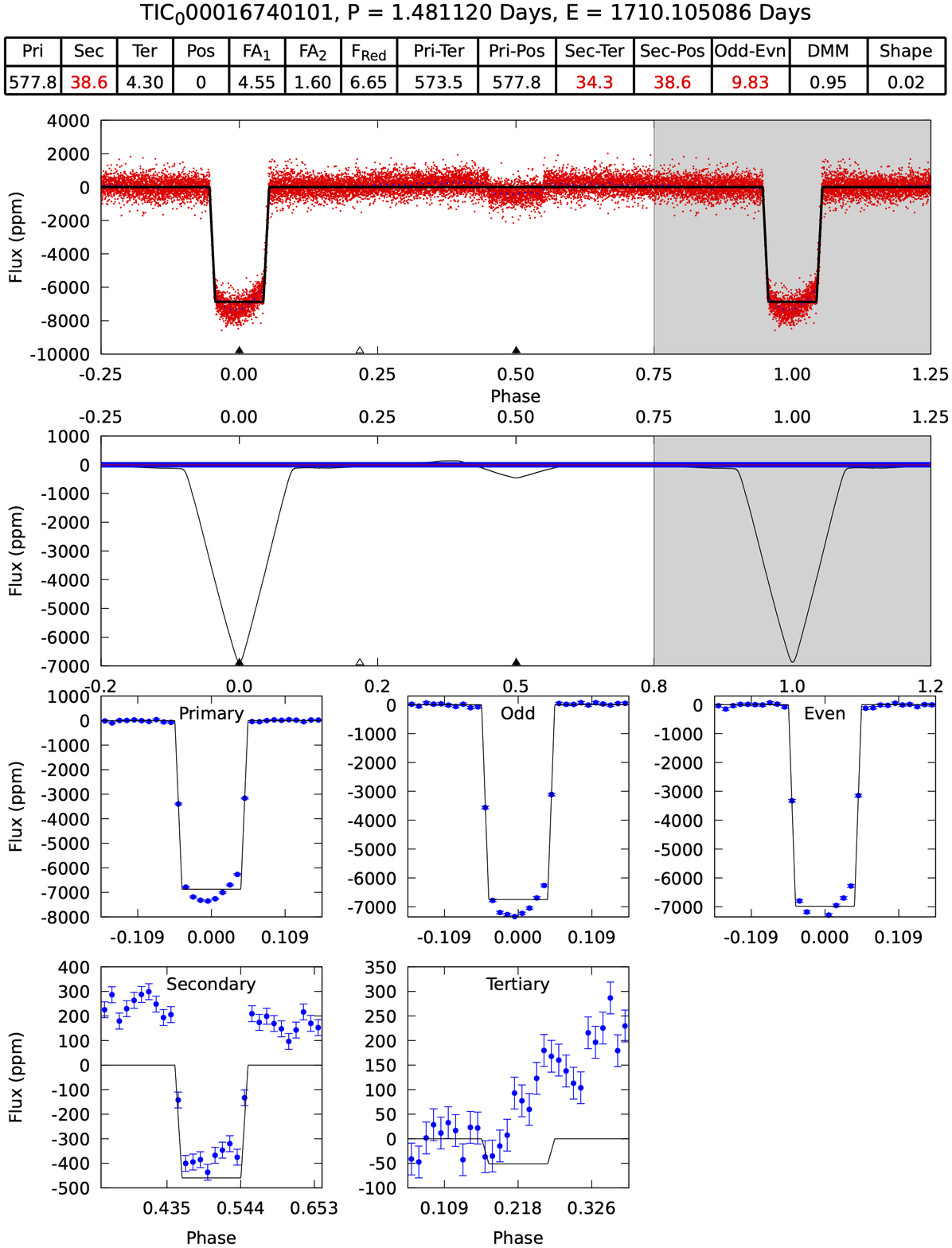}
    \caption{The phase-folded light curve for TIC 16740101, also known as KELT-9 b. A secondary eclipse on the same period of the input signal is clear in \dave Modelshift result. The secondary is due to the occultation of the bright side of the planet when it orbits behind the star along TESS line of sight. Based on \dave results only, for consistency we label this TOI as FP due to the significant secondary eclipses.}
    \label{fig:occ}
\end{figure*}

\bsp	
\label{lastpage}

\end{document}